\documentclass[aps,prb,twocolumn,superscriptaddress,floatfix]{revtex4-1}

\usepackage{amssymb,amsmath,slashed,comment}
\usepackage{hyperref}
\usepackage{color}
\usepackage{graphicx}
\usepackage{comment}
\usepackage{enumitem}

\begin{document}

\title{Boundary conformal field theory and symmetry-protected topological phases in 2 + 1 dimensions}

\author{Bo Han} 
\affiliation{
Department of Physics, University of Illinois
at Urbana-Champaign, 
1110 West Green Street, Urbana, Illinois 61801, USA
            }

\author{Apoorv Tiwari}
\affiliation{
Department of Physics, University of Illinois
at Urbana-Champaign, 
1110 West Green Street, Urbana, Illinois 61801, USA
            }

\author{Chang-Tse Hsieh}
\affiliation{
Department of Physics, University of Illinois
at Urbana-Champaign, 
1110 West Green Street, Urbana, Illinois 61801, USA
            }

\author{Shinsei Ryu}
\affiliation{
  James Franck Institute and Kadanoff Center for Theoretical Physics,
  University of Chicago, Illinois 60637, USA}

\date{\today}

\begin{abstract}
We propose a diagnostic tool for detecting nontrivial symmetry-protected topological (SPT) phases 
protected by a symmetry group $G$ in 2 + 1 dimensions. 
Our method is based on directly studying the 1 + 1-dimensional anomalous edge conformal field theory (CFT) of 
SPT phases. 
We claim that if the CFT is the edge theory of an SPT phase, then there must be an obstruction to cutting it open. 
This obstruction manifests as the in-existence of boundary states in the CFT
that preserves both the conformal symmetry and the global symmetry $G$. 
We discuss the relation between edgeability, i.e., the ability to find a consistent boundary state, 
and gappability, i.e., the ability to gap out a CFT, in the presence of $G$. 
We study several cases including time-reversal symmetric topological insulators,
$\mathbb{Z}_N$ symmetric bosonic SPT phases,
and $\mathbb{Z}_2 \times \mathbb{Z}_2$ symmetric topological superconductors. 
\end{abstract}


\maketitle

\tableofcontents

\section{Introduction}

\subsection{SPT phases and quantum anomalies}

Symmetry-protected topological (SPT) phases are quantum states of matter with a
global symmetry $G$,
which can be either an internal or a space-time symmetry \cite{gu2009tensor,Chen2011a,Chen2011b,Hasan2010a,Qi2011a,Fu2011a,Song2017a,2016RvMP...88c5005C}.
This symmetry prevents one from adiabatically connecting an SPT state to a topologically trivial state, 
namely, a product state. 
More precisely, this means that one cannot find a symmetry-preserving quasilocal unitary transformation that maps an SPT state to a product state
\cite{gu2009tensor}.
In fact,
as long as the symmetry is unbroken (either explicitly or spontaneously),
the phase space of gapped systems is partitioned into topologically distinct sectors 
that cannot be adiabatically connected to each other.
The trivial state lies in the trivial (in the topological sense) sector of this classification.
We will henceforth refer to it as the trivial SPT phase. 

The existence of SPT phases has a close connection with quantum anomalies which are purely quantum phenomena without any classical analog. 
Crudely speaking, 
it is expected that on the $d$-dimensional boundary of 
an SPT phase in $d+1$-dimensional space-time 
lives an interesting phase of matter 
which is anomalous in the sense that it cannot exist 
on a pure $d$-dimensional spacetime manifold,
but must always be realized on the boundary of a
$d+1$ manifold under the condition that the symmetry is realized in the same way as
in the bulk
\cite{LuVishwanath2012,chen2012symmetry,chen2013symmetry,pollmann2012symmetry,fidkowski2011topological}
.

Put differently,
consistency conditions of 
a conformal field theory (CFT) at the boundary and 
topological properties of a bulk phase
are closely related. 
Basically, a consistent CFT, when realized at the edge of a bulk system, 
implies that the bulk is trivial, namely, continuously deformable to a trivial state. 
On the other hand, if a bulk supports a CFT that is inconsistent as its boundary theory, the bulk cannot be deformed into a trivial state. 
This leads us to the question: \emph{What are the criteria for a CFT to be consistent? }

For a (2+1)$d$ SPT phase, 
modular (non)invariance of its (1+1)$d$ edge theory has been used as a diagnostic for the (non)trivial bulk 
\cite{ryu2012interacting,sule2013symmetry}. 
For a nontrivial SPT phase protected by symmetry $G$, there is a conflict between $G$ and modular invariance of the edge CFT; more precisely, the edge theory orbifolded by $G$ is not modular invariant\cite{ryu2012interacting,sule2013symmetry}.
A similar argument can be applied to boundary theories of SPT phases in space dimensions higher than 2
\cite{hsieh2016global, witten2016fermion}.

\subsection{Edgeability}

In this paper, we will give further thought on consistency conditions of CFTs. 
We will rely on the simple geometrical identity, 
$\partial^2 = 0$, 
which essentially says there is no boundary to a boundary of a bulk system. 
This would mean, in the context of SPT phases and their boundaries, 
that boundary theories of SPT phases are not allowed to have boundaries.
Conversely, it is likely that 
any ``healthy'' (conformal) field theories should be possible to have boundaries $-$
this may be a consistency condition of the (conformal) field theories. 

In studies of surface topological orders of (3 + 1)$d$ SPT phases,
such a consistency condition was called ``edgeability'' 
\cite{vishwanath2013physics, Chen201509a}.
(2 + 1)-dimensional surface theories are said to be ``nonedgeable'',
meaning that it is not possible to create an edge between the theory in question and the vacuum. 
The only boundary that one can possibly create is a domain wall. 
In contrast, any consistent (2 + 1)$d$ theory should be ``edgeable'' to the trivial vacuum.
Here ``edgeability'' may be also called ``cuttability'',
meaning that the original theory defined on a closed spacetime 
can be cut open.

We will follow this idea but focus on one lower dimension.  
In (1 + 1)$d$ CFTs, 
in addition to modular invariance, it is often claimed that a consistent conformal field theory with boundaries must have a complete set of boundary states.
(See, for example, Refs.\ \onlinecite{2000NuPhB.570..525B,sousa2003orientation}.)
This reminds us of edgeability. 
In fact, the construction of modular invariant partition functions are closely related to boundary conformal field theories (BCFTs). 
The perspective from SPT phases gives us some insight about
why BCFT is crucial for consistency,
which may look a little puzzling from other viewpoints.
In order for a CFT to exist as a pure (1 + 1)$d$ system, 
both edgeability and modular invariance must be satisfied.
(Once again, here edgeability may be also called cuttability,
meaning that the original closed (1 + 1)$d$ CFT
can be cut open into a well-defined BCFT.)
In many known cases, 
these two conditions are actually equivalent.

The simplest examples of (non)edgeability
can be provided by chiral edge theories of topologically ordered
phases, in which the chiral central charge is nonzero.
In these edge theories, it is not possible to find conformally invariant
boundary conditions or boundary states. 
These conformal field theories 
are hence non-edgeable and must be realized at 
the boundary of a (topological) phase in one higher dimensions. 

In the context of SPT phases, 
Ref.\ \onlinecite{santos2014symmetry}, 
gives an explicit lattice construction of 
an (1 + 1)$d$ CFT,
which is the edge theory of (2 + 1)$d$ bosonic SPT phases
protected by $\mathbb{Z}_N$ on-site unitary symmetry.
In this construction, while the CFT is successfully put on a one-dimensional lattice,  
the $\mathbb{Z}_N$ symmetry is not realized 
in a purely local way -- 
the action of the $\mathbb{Z}_N$ transformation is non-on-site,
and it involves links of the lattice. 
It was claimed that its non-on-site symmetry has been gauged,
which is equivalent to orbifolding $\mathbb{Z}_N$ symmetry. 
As we will clarify, 
within this CFT with the non-onsite action of the $\mathbb{Z}_N$ symmetry
(and its lattice realization), 
it is not possible to make a boundary which is consistent
with the ${\mathbb{Z}}_N$ symmetry. 
Hence, this theory is nonedgeable.

\begin{figure}
\hspace{0cm}  \vspace{0cm}
\includegraphics[width=8cm,height=3cm]{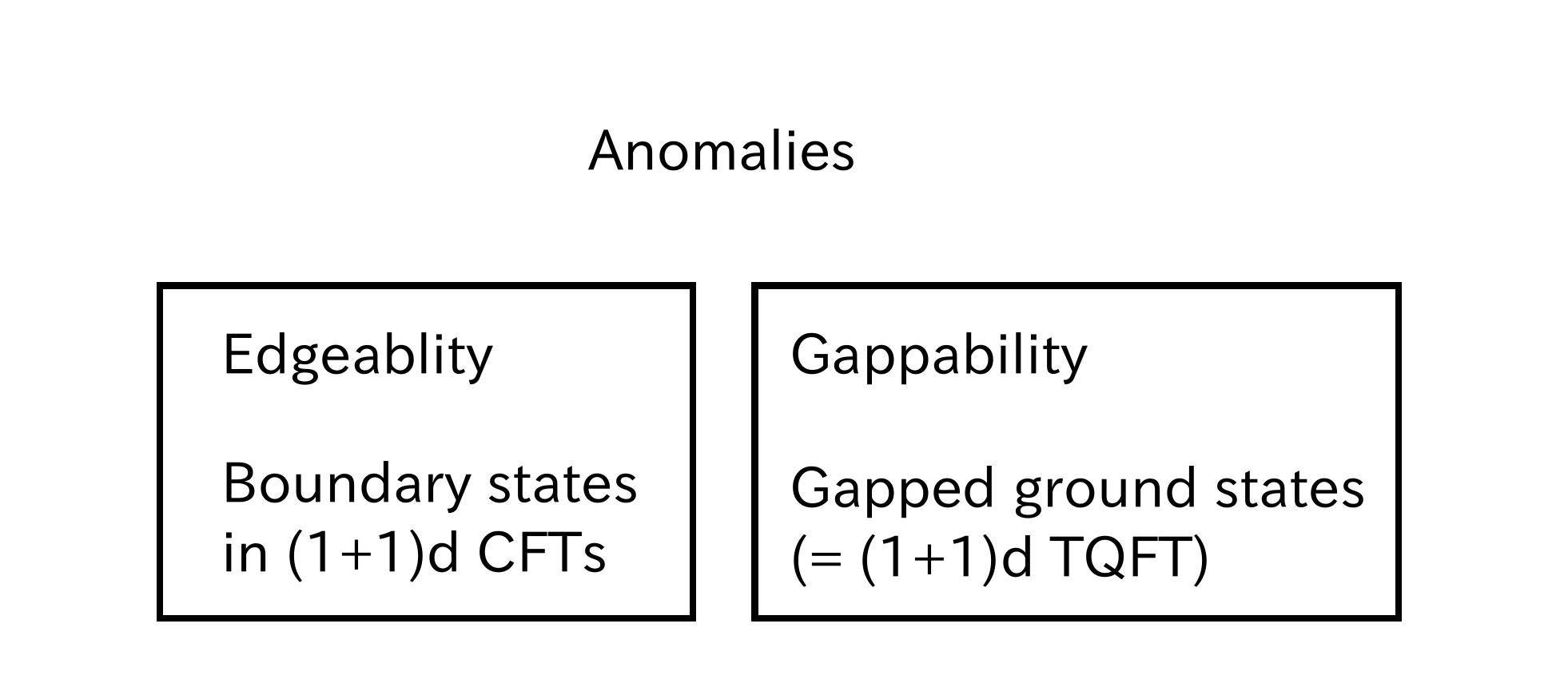}
\caption{
Edgeability and gappability of conformal field theories are closely related -- they both diagnose if they must be realized 
as a boundary of (topological) systems in one higher dimensions.
Hence, edgeability and gappability are both related to quantum anomalies.   
}
\label{anomaly}
\end{figure}  

\subsection{Gappability}

Let us now also give a slightly different perspective
from the correspondence between (1 + 1)$d$ gapped states 
and boundary states in CFTs.
In Refs.\ \onlinecite{2016arXiv160606402C,2015JHEP...05..152M},
(1 + 1)$d$ conformal field theories perturbed by 
some operators are considered.
If the perturbation is such that it fully gaps out the theory, 
we flow from the CFT to a massive phase. 
It was claimed that the ground state of the massive phase
is described, with the Hilbert space of the CFT, by
a boundary state.
In particular, in Ref.\ \onlinecite{2016arXiv160606402C}
this claim is explicitly verified 
for various symmetry-protected topological phases in (1 + 1)$d$,
which are obtained by perturbing CFTs.
These phases are fully gapped (1 + 1)$d$ phases protected by a certain set of symmetries.
In particular, topological invariants, for instance, the group cocycle $\varepsilon \in H^2(G, \text{U(1)})$ 
of the group cohomology classification of (1 + 1)$d$ SPT phases
\cite{chen2012symmetry,chen2013symmetry},
can be fully extracted from boundary states that describe SPT phases.

In this paper, instead of (1 + 1)$d$ SPT phases, we are concerned with
(2 + 1)$d$ SPT phases, and in particular their edge theories. 
In various examples, 
we establish a claim similar to the above claim for the bulk (1 + 1) dimensions;
we again establish a connection between 
gapped ground states in the edge theories 
and 
conformal boundary states.
The relation between edgeability and gappability is schematically
illustrated in Fig.\ \ref{anomaly}.
In particular, 
for the edge theories described by the $K$-matrix theory,
we establish the connection between
Haldane's null vector criterion for gapping potentials
\cite{1995PhRvL..74.2090H}
and the boundary states. 

More precisely,
the main question we ask is which boundary conditions (including symmetry projections) can be imposed on conformal field theories that are defined on the edge of (2 + 1)$d$ systems with boundaries.  
We show the equivalence between our BCFT formalism and the $K$-matrix formalism
used in Ref.\ \onlinecite{LuVishwanath2012}
and show that those CFTs that admit a consistent boundary state correspond to edge theories of trivial SPTs and those that do not admit a consistent boundary state correspond to edges of nontrivial SPTs.

This criterion is very similar to that imposed by the modular invariance
of CFTs field theories on the torus. 
Putting a theory on a torus in this context implies that it can be realized on a strictly (1 + 1)$d$ manifold and need not be realized on the boundary of an SPT phase{\cite{ryu2012interacting,sule2013symmetry, cho2015topological,hsieh2014symmetry}}.

\subsection{Working Principles}

\begin{figure}
\hspace{0cm}  \vspace{0cm}
\includegraphics[width=8cm,height=3cm]{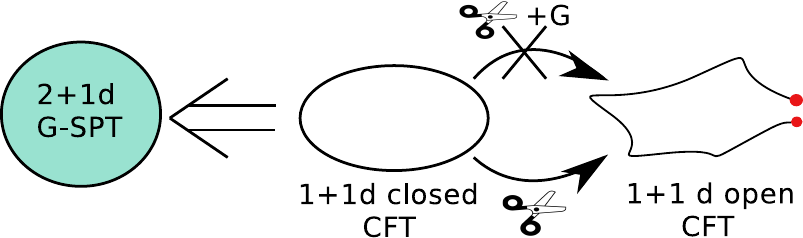}
\caption{We claim that one cannot ``cut'' or ``make a boundary''
  while preserving $G$ symmetry for certain CFTs and certain symmetry
  implementations. These symmetry implementations correspond precisely to
  $(2 + 1)$d $G$-symmetric SPT phases and the corresponding CFTs are their edge theories.}
\label{cartoon}
\end{figure}  

Following these motivations,
let us now describe our  
strategy to detect and diagnose non-trivial SPT phases. 
We claim that one cannot construct a symmetry invariant Cardy state (conformal boundary state) for a CFT corresponding to the edge of a non-trivial SPT. 
As mentioned in the above example,
this is due to the fact that
although one may be able to put the edge theory of an SPT
on a lattice,
$G$ symmetry cannot be implemented in an on-site way --
this shows up as nonedgeablity.   

A Cardy state in conformal field theory is a coherent state 
in the Hilbert space of the closed sector of the CFT which satisfies an open-closed consistency relation,
namely the Cardy condition \cite{cardy1989boundary,cardy2004boundary}.
We show that in the case of nontrivial SPT phases, it is not possible to implement the symmetry and simultaneously satisfy the Cardy condition. 

We list our procedure for diagnosis of SPT phases as follows
(see Fig.\ \ref{cartoon} for an illustration).  
\begin{enumerate}[label=(\roman*)]
\item First, cut the 1$d$ circle and impose appropriate boundary conditions on the two ends.
\item Second, solve for Ishibashi states \cite{ishibashi1989boundarycrosscap} of this open system, which correspond to solutions to the boundary conditions imposed.
\item Third, try to construct a boundary state that is a linear combination of
  Ishibashi states, which satisfies the Cardy condition and is also symmetry
  invariant. If such a state exists, then there is no nontrivial SPT phase in
  a (2 + 1)$d$ bulk system; if such a state does not exist,
  then the corresponding (2 + 1)$d$ bulk is a nontrivial SPT phase. 
\item Fourth, once we have detected nontrivial SPT phases, we can obtain its classification from the transformation of the Cardy state under symmetry operation, which will be explained in detail in later sections. 
\end{enumerate}
Using this technique, we can study SPT phases protected by space-time and/or some
internal symmetries.
Examples include the time-reversal symmetric topological insulators, bosonic SPT
phases with $\mathbb{Z}_N$ symmetry, and topological superconductors protected
by $\mathbb{Z}_2 \times \mathbb{Z}_2$ symmetries.
These examples have been also analyzed in the literature
by different methods
\cite{LiuZX2013a,WenXG2013a,ChenX2011b,Levin2012a,Yao2013a}.
 
The organization of the rest of the paper is as follows. 
In Sec.\ \ref{sec:bcftIntro}, a brief introduction to BCFT is provided. 
In Sec.\ \ref{Edge theories of (2+1)d time-reversal symmetric topological insulators}, we study (2 + 1)$d$ time-reversal symmetric topological insulators from the edge theories and the corresponding symmetric Cardy boundary states.  
Edge theories of more general (2 + 1)$d$ SPT phases described
by the $K$-matrix theories are considered in Sec.\ \ref{More general SPT phases in (2+1)D}, where a connection between the Cardy states and gapped phases in (1 + 1) dimensions is shown explicitly.
Then we apply our approach to topological superconductors in Sec.\ \ref{(2+1)D topological superconductor protected by Z2xZ2 symmetry}.
Discussions and conclusions are given in Sec. \ref{sec:Discussion}.

\section{An introduction to BCFT}
\label{sec:bcftIntro}

A boundary condition in a CFT defines a relation between 
the holomorphic and antiholomorphic sectors. 
In other words, 
the two sectors are related to one another on the boundary 
via an automorphism of the form
\begin{align}
S(z)=\rho_{\beta}\left(\bar{S}(\bar{z})\right), 
\end{align}   
where 
$S$ belongs to some symmetry algebra,
$\rho_{\beta}$ denotes an automorphism of the algebra of fields,
and $\beta$ is a constant that parametrizes the boundary condition.
$S(z)$ and $\bar{S}(\bar{z})$ are fields which have the following expansion in terms of modes:
\begin{align}
S(z)=\sum_{n\in\mathbb Z}{S_{n}z^{-n-h}}, \quad 
\bar{S}(\bar{z})=\sum_{n\in\mathbb Z}{\bar{S}_{n}\bar{z}^{-n-\bar{h}}},
\end{align}
where $h$ and $\bar{h}$  are the conformal weights of $S$ and $\bar{S}$ respectively. 
In most general situations,
$S$ and $\bar{S}$
are the holomorphic and antiholomorphic components of
the stress-energy tensor with $h=2$.
For CFTs with current algebra symmetries,
$S$ and $\bar{S}$
are taken to be the currents with $h=1$.

\begin{figure}
\label{open-closed_corresp}
\hspace{-4cm}  \vspace{0cm}
\includegraphics[width=12cm,height=5.5cm]{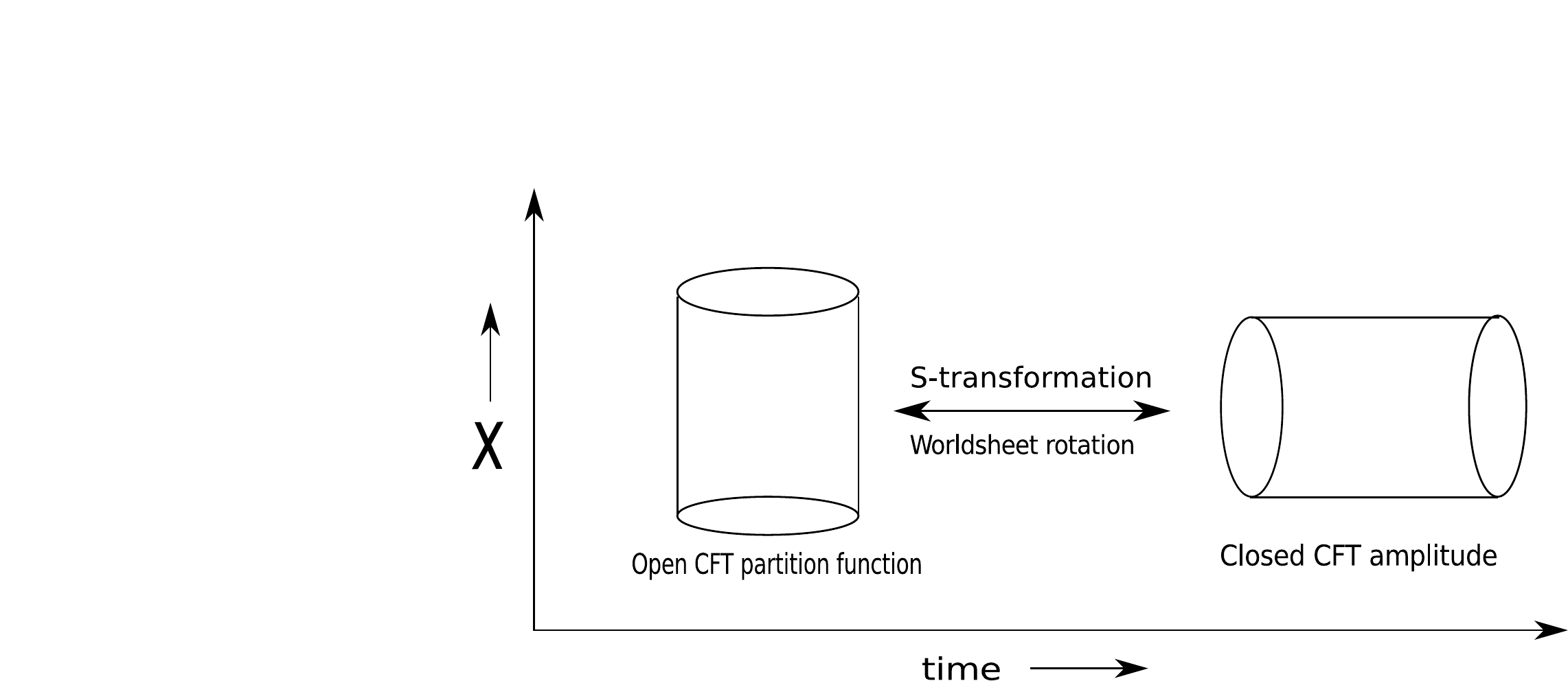}
\caption{An illustration of the Cardy condition; a consistency condition for conformal boundary states. For boundary states that preserve conformal symmetry, the open channel partition function $ Z_{\text{open}}:=\text{Tr}_{\mathcal H_{\text{open}}}\left[e^{-(2\pi i/\tau) H_{\text{open}}}\right]$ must equal the amplitude for a ``Cardy'' state $\mathcal A=\langle B|e^{2\pi i \tau H_{\text{closed}}}|B\rangle$}.
\label{worldsheetrotation}
\end{figure}

In the closed picture, a boundary condition is represented 
by a state in the Hilbert space of a CFT defined on a circle.
According to Cardy, 
such a boundary state must transform consistently under 
the $S$-modular transformation, namely, a $\pi/2$ rotation of the space-time manifold (worldsheet) illustrated in Fig. \ref{worldsheetrotation}.
To construct physical boundary states obeying this 
consistency condition (the Cardy condition),
one first constructs a set of states,
the so-called Ishibashi states,
$|i,\beta\rangle\!\rangle$, 
which are annihilated by the boundary conditions (known as gluing conditions) in the operator form  after the ${\pi}/{2}$ rotation,  
$\left[S_n-\rho_{\beta}\left(\bar{S}_{n}\right)\right]
\xrightarrow{\text{worldsheet  rotation}}
\left[S_n-(-1)^{h}\rho_{\beta}\left(\bar{S}_{n}\right)\right]$,
namely, 
\begin{align}
\left[S_n-(-1)^{h}\rho\left(\bar{S}_{n}\right)
\right]|i,\beta\rangle\!\rangle =0.
\end{align}
A Cardy state is a suitable superposition of the Ishibashi states that satisfy the Cardy condition, which is an implementation of open-closed channel consistency: 
\begin{align}
Z_{\alpha\beta}(-1/\tau )&=\langle B_{\alpha} |e^{2\pi i \tau H_{\text{closed}}}|B_{\beta}\rangle,
\label{Cardycondition}
\end{align} 
where $Z_{\alpha\beta}$ is 
the partition function computed in the open-channel picture,
and given as a trace over the open Hilbert space with boundary conditions
$\alpha,\beta$ on the two ends,
and $\tau$ is parameterized by the size of the system \cite{gaberdielnotes}.
The states $|B_{\alpha}\rangle$ and $|B_{\beta}\rangle$ which satisfy the above condition are the bona fide boundary states,
the Cardy states. 
For a more detailed introduction to boundary conformal field theory, see, 
for example, 
Refs. \onlinecite{cardy1989boundary,cardy2004boundary,FMS-CFT,gaberdielnotes,blumenhagen2009introduction,recknagel2013boundary}.

\paragraph*{Symmetry invariant Cardy states and the obstruction}

A generic Cardy boundary state $|B\rangle$ 
(here we are suppressing the label $\alpha,\beta$ specifying boundary conditions) 
lies in the subspace of the closed Hilbert space and satisfies
\begin{align}
\label{BSs_in_T}
(T-\bar{T})|B\rangle=0 \quad \text{(on a boundary)},
\end{align}
where $T$ and $\bar{T}$ are the holomorphic and anti-holomorphic parts of the energy density operator, 
respectively. 
In terms of the Virasoro generators, 
this condition can be written as 
\begin{align}
(L_n -\bar{L}_{-n})|B\rangle &= 0. \label{eqn:VirasoroBdyState}
\end{align}

Equation \eqref{eqn:VirasoroBdyState}
implies $(c-\bar{c})|B\rangle =0$, where $c$ and $\bar{c}$ represent the central
charges for the holomorphic and antiholomorphic sectors, respectively.
Thus, as expected, one cannot construct (conformally invariant) boundary states
when $c\neq \bar{c}$ since in this case, the (1 + 1)d CFT suffers from the
(infinitesimal) gravitational anomaly, and hence it must be realized as
the boundary theory of an appropriate bulk system living in one higher
dimension. 

%

For the rest of the paper, we will deal with systems with the vanishing chiral
central charge, $c-\bar{c}=0$, and hence there is no infinitesimal gravitational
anomaly.
We will also focus on (1 + 1)$d$ CFTs for which one can construct a modular
invariant, if one is willing not to impose any additional symmetry. 
Hence, in the absence of symmetries, the (1 + 1)$d$ CFT can be safely gapped
by adding a suitable perturbation. 
However, if we impose some symmetry, e.g., if we consider (1 + 1)$d$ CFTs realized
potentially on the boundary of (2 + 1)$d$ SPT phases,
there may be a conflict between the symmetry and modular invariance. 
Once symmetry is gauged (orbifold), the modular invariance may be spoiled. 
Conversely, if the modular invariance is enforced, the symmetry must be broken.


At the level of BCFT, this would mean that
one may not be able to construct boundary states which are invariant under the symmetry.
More precisely, 
we consider a symmetry that 
preserves $T$ and $\bar{T}$, respectively, 
or exchanges them. 
Classically, such a symmetry preserves the conformally invariant boundary condition $T=\bar{T}$ along the boundary. 
However, once the theory is quantized, there may be an 
obstruction to construct the corresponding 
a boundary state.
A symmetry of a CFT (on a closed space-time manifold) is anomalous
if one cannot make a boundary that preserves this symmetry both classically and quantum mechanically. Typically, this happens when there is a conflict between the Cardy condition (\ref{Cardycondition}) and the symmetry, so a symmetric Cardy state does not exist. 
In this situation, the theory itself, together with the symmetry,
cannot be consistently defined and must appear as a boundary theory of a SPT
phase with the same symmetry in one higher dimensions.
Nevertheless, by ``stacking'' copies of such SPT phases,
the number of degrees of freedom at boundaries increase,
and the solution space of Eq. (\ref{BSs_in_T}) is enlarged -- 
it may be possible to find
a symmetric Cardy state if the number of the copies
is large enough.
When this occurs,
the corresponding CFT is anomaly free with respect to such a symmetry and can exist alone in its own dimension.

\section{Edge theories of (2 + 1)$d$ time-reversal symmetric topological insulators}
\label{Edge theories of (2+1)d time-reversal symmetric topological insulators}
Let us begin with a simple example.
Consider the edge theory of a (2 + 1)$d$ time-reversal symmetric topological insulator, which is described by (1 + 1)$d$ massless Dirac fermions on a closed two-manifold $\Sigma$:
\begin{align}
\label{massless_Dirac}
S=\frac{1}{2\pi} \int_\Sigma 
{d}t{d}x
\left(i\bar{\psi}_{R}\partial_{+}{\psi}_{R}+i\bar{\psi}_{L}\partial_{-}{\psi}_{L}\right),
\end{align}
where $\partial_{\pm}=\partial_{t}\pm\partial_x$. 
The system is invariant under both charge U(1)$_C$ and time-reversal symmetries, which are defined as  
\begin{align}
\text{U(1)}_{C}:&\; \psi_{R}\to e^{-i\theta}\psi_{R}, 
\quad \psi_{L}\to e^{-i\theta}\psi_{L}
\nonumber \\
\mathcal{T}_{\eta}:&\; \psi_{R}\to \psi_{L}, 
\quad \psi_{L}\to \eta\psi_{R}, \quad \eta=\pm 1.
\end{align}
Here in principle we have two choices for time-reversal symmetry (characterized by $\eta$): $\mathcal{T}_{1}^2=1$  and $\mathcal{T}_{-1}^2=(-1)^F$, where $F$ is the total fermion number operator.
By analyzing the stability (gappability) of the theory (\ref{massless_Dirac}), at least at the quadratic level (namely, by considering adding symmetry-respecting fermion mass bilinears to the action), we know that $\eta=1$ ($\eta=-1$) corresponds to the edge of the topologically trivial (nontrivial) phase. It can also be shown the nontrivial topological phases form a $\mathbb{Z}_2$ class.

Now let us study the same problem (classification of topological insulators) by the BCFT approach.
Consider putting the theory (\ref{massless_Dirac}) on a cylinder $\Sigma$ with boundary at $x=0,\pi$. 
Then we would like to know, based on the discussion in the previous sections, if there exists a Cardy boundary state, which satisfies the conditions (\ref{Cardycondition}) and (\ref{BSs_in_T}), invariant under U(1)$_C$ and $\mathcal{T}_\eta$. If such a Cardy state does not exist, the corresponding theory must be the edge of a (2 + 1)$d$ topological insulator.

One obtains the boundary conditions by varying the action (\ref{massless_Dirac}) on the cylinder $\Sigma$,
\begin{align}
\left.\left[\bar{\psi}_{R}\delta \psi_{R} +\psi_{R}\delta \bar{\psi}_{R}
-\bar{\psi}_{L}\delta \psi_{L}-\psi_{L}\delta \bar{\psi}_{L}\right]\right|_{\partial\Sigma}=0.
\end{align}
This boundary condition can be solved by the following set of gluing conditions \cite{Hori2003a}:
\begin{align} 
B_{\beta}\text{ type}: \quad\psi_{L}=&\; e^{-i\beta}\psi_{R}, \quad \bar{\psi}_{L}=e^{i\beta}\bar{\psi}_{R}, \nonumber \\
A_{\alpha}\text{ type}: \quad\psi_{L}=&\; e^{-i\alpha}\bar{\psi}_{R}, \quad \bar{\psi}_{L}=e^{i\alpha}{\psi}_{R} .
\label{complexfermionbc}
\end{align}
The two kinds of boundary conditions have been labeled $B_{\beta}$ and $A_{\alpha}$, respectively.
Note that, as the bulk theory respects all symmetries, the presence of boundary might in general break (some of) the symmetries. To be specific, the $B_\beta$ boundary condition (with arbitrary $\beta$) preserves both U(1)$_C$ and $\mathcal{T}_{1}$, while either the $B_\beta$ or the $A_\alpha$ boundary condition cannot preserve both U(1)$_C$ and $\mathcal{T}_{-1}$. 
Therefore, it is impossible to, at least for a single copy of the theory (\ref{massless_Dirac}), find a symmetric Cardy state with respect to both U(1)$_C$ and $\mathcal{T}_{-1}$, because there is no such symmetry invariant boundary condition.

Let us first focus on the case of the symmetry group U(1)$_C\rtimes\mathbb{Z}_2^{\mathcal{T}_1}$, where $\mathbb{Z}_2^{\mathcal{T}_1}$ is generated by $\mathcal{T}_1$. Although the $B_\beta$ boundary condition preserves U(1)$_C\rtimes\mathbb{Z}_2^{\mathcal{T}_1}$, we still have to check that the corresponding Cardy state is also symmetry invariant.

We impose boundary conditions 
$B_0, B_\beta$ at $x=0,\pi$, respectively. 
In order to satisfy the boundary conditions, we define a mode expansion
\begin{align}
\psi_{L}=&\;\sum_{r\in \mathbb{Z}+\beta/2\pi}{\psi_{r}(t)e^{irx}}, \quad \bar{\psi}_{L}=\sum_{r''\in \mathbb{Z}-\beta/2\pi}{\bar{\psi}_{r'}(t)e^{ir'x}}, \nonumber \\ 
\psi_{R}=&\;\sum_{r\in \mathbb{Z}+\beta/2\pi}{\psi_{r}(t)e^{-irx}}, \quad \bar{\psi}_{R}=\sum_{r'\in \mathbb{Z}-\beta/2\pi}{\bar{\psi}_{r'}(t)e^{-ir'x}}.
\end{align}
The mode operators satisfy the following algebra:
\begin{align}
&
\{\psi_{r}(t),\psi^{\dagger}_{r'}(t)\}=2\pi\delta_{r+r',0}, 
\nonumber  \\
&
\{\psi_{r}(t),\psi_{r'}(t)\}=
\{\psi^{\dagger}_{r}(t),\psi^{\dagger}_{r'}(t)\}=0.
\label{complexfermionalgebra}
\end{align}
We define the normal ordering with respect to a vacuum $|0,\beta\rangle$ which is annihilated by $\psi_r$ ($\geq 0$) and $\bar{\psi}_{r'}$ ($\geq 0$):
\begin{align}
:\bar{\psi}_{-r}\psi_{r}:=
\begin{cases}   
\bar{\psi}_{-r}\psi_{r} \quad \mbox{if} \ r\geq 0 ,  \\
-\psi_{r}\bar{\psi}_{-r} \quad \mbox{if} \ r<0.
\end{cases}
\end{align}
The Hamiltonian and U(1)$_C$ charge operator $F$ take the form
\begin{align}
H_{o}
=&\; \sum_{r\in \mathbb Z+\frac{\beta}{2\pi}}{r:\bar{\psi}_{-r}\psi_{r}:+\frac{1}{2}\left(\frac{\beta}{2\pi}-\left[\frac{\beta}{2\pi}\right]-\frac{1}{2}\right)^2-\frac{1}{24}}, \nonumber \\
F=&\; \sum_{r\in \mathbb Z+\frac{\beta}{2\pi}}{:\bar{\psi}_{-r}\psi_{r}:+\frac{\beta}{2\pi}-\left[\frac{\beta}{2\pi}\right]-\frac{1}{2}}.
\end{align}
The open-channel partition function with insertion of symmetry flux $e^{-2\pi i(a-1/2)F}$ on the cylinder with $(\ell_{\mathrm{space}}, \beta_{\mathrm{time}})=(\pi, 2\pi T)$ is 
\begin{align}
Z_{0\beta}^a(T)&= \mathrm{Tr}_{\mathcal{H}_{o}}\left[e^{-2\pi i(a-1/2)F}e^{-2\pi TH_{o}}\right] 
\nonumber\\
&=\frac{\vartheta
\left[
 \begin{smallmatrix}
\beta/2\pi - 1/2 \\ -(a-1/2)
\end{smallmatrix}
\right]
(0,iT)}{\eta(iT)}.
\label{openbeta}
\end{align}

To construct the Cardy states, we work in Euclidean signature, by performing the Wick rotation $t=-i\tau$, and consider boundary conditions in the closed channel [after the space-time cylinder has been rotated by $\pi/2$, namely, $(x', \tau') = (\tau, -x)$]:
\begin{align}
&\psi'_{L}= e^{-i\beta}e^{-i\pi/2}\psi'_{R}, \quad
\bar{\psi}'_{L}=e^{i\beta}e^{-i\pi/2}\bar{\psi}'_{R}, 
\label{betagluingcond_after_rotation}
\end{align}
where we have introduced the notation
\begin{align}
\label{fermions_after rotation}
\psi'_{R}&=e^{i\pi/4}\psi_R, \quad\bar{\psi}'_{R}=e^{i\pi/4}\bar{\psi}_R, \nonumber \\
\psi'_{L}&=e^{-i\pi/4}\psi_L, \quad\bar{\psi}'_{L}=e^{-i\pi/4}\bar{\psi}_L,
\end{align}
for the fields with respect to the coordinate system after the $\pi/2$ space-time rotation.

In Euclidean signature, the original time-reversal symmetry
$\mathcal{T}_{\eta}$, which is an anti-unitary operator in the Lorentz signature, becomes the unitary $(\mathcal{CP})_{\eta}$ symmetry, the product of charge conjugation and spatial reflection that flips $\tau$ to $-\tau$. From the relation (\ref{fermions_after rotation}), this $(\mathcal{CP})_{\eta}$ is further translated to $(\mathcal{CP})'_{-\eta}$, which acts on the fermions as
\begin{align}
\label{CP'_after_rotation}
(\mathcal{CP})'_{-\eta}:\ &\psi_{R}'(x', \tau')\to e^{i\pi/2}\bar{\psi}'_{L}(-x', \tau'), 
\nonumber\\
&\psi'_{L}(x', \tau')\to \eta e^{-i\pi/2}\bar{\psi}'_{R}(-x', \tau').
\end{align}
Therefore, under the $\pi/2$ space-time rotation (together with an analytic
continuation from the Lorentz to Euclidean signature), we have the following correspondence:
\begin{align}
\mathcal{T}_{\eta}^2= \eta^{F} \longleftrightarrow {(\mathcal{CP})'}_{-\eta}^2= (-\eta)^{F}, \quad \eta=\pm1.
\end{align}
One can check that $(\mathcal{CP})'_{-1}$ preserves the boundary condition (\ref{betagluingcond_after_rotation}) as $\mathcal{T}_{1}$ preserves the $B_\beta$-type boundary condition in (\ref{complexfermionbc}).

Now, since we inserted a U(1)$_C$ charge operator in the trace when evaluating the open-channel partition function, the corresponding boundary states in the closed channel must lie in the subspace of the Hilbert space of the twisted fields that satisfy 
\begin{align}
\psi_{R}'(x'+2\pi,\tau')=&\; e^{2\pi ia}\psi_{R}'(x',\tau'), \nonumber \\
\psi_{L}'(x'+2\pi,\tau')=&\; e^{2\pi ia}\psi_{L}'(x',\tau').
\label{WSBC}
\end{align}     
(Here we compactify the space direction as $x'\equiv x'+2\pi$.)
Hence, we get the following mode expansions:
\begin{align}
\label{mode_after_rotation}
\psi_{R}'=&\; \sum_{r\in \mathbb Z+a}{\psi'_{r}e^{irw'}}, \quad \bar{\psi}_{R}'= \sum_{r'\in \mathbb Z-a}{\bar{\psi}'_{r'}e^{ir'w'}}, \nonumber \\
\psi_{L}'=&\; \sum_{\tilde{r}\in \mathbb Z-a}{\tilde{\psi}'_{\tilde{r}}e^{-i\tilde{r}w'}}, \quad \bar{\psi}_{L}'= \sum_{\tilde{r}'\in \mathbb Z+a}{\bar{\tilde{\psi}}_{\tilde{r}'}'e^{-i\tilde{r}'w'}},
\end{align}
where $\omega'=x'+i\tau'$. 
The Hamiltonian is
\begin{align}
H_{c}
&= \sum_{r\in \mathbb Z+a}r:\bar{\psi}_{-r}'\psi_{r}':
+\sum_{\tilde{r}\in \mathbb Z-a}\tilde{r}:\bar{\psi}_{-\tilde{r}}'\psi_{\tilde{r}}':
\nonumber\\
&+\frac{1}{2}\left(a-[a]-\frac{1}{2}\right)^2+\frac{1}{2}\left(a+[-a]+\frac{1}{2}\right)^2-\frac{1}{12}.
\end{align}
The ground state $|0\rangle_{a,-a}$ is defined to be the state annihilated by $\psi'_r\ (r \geq 0),\ \bar{\psi}'_{r'}\ (r' > 0),\ \tilde{\psi}'_{\tilde{r}}\ (\tilde{r} \geq 0),\ \bar{\tilde{\psi}}'_{\tilde{r}'}\ (\tilde{r}' > 0)$.

The gluing condition (\ref{betagluingcond_after_rotation}) for the mode operators takes the form\begin{align}
\psi_{r}'=&\; ie^{i\beta}\tilde{\psi}_{-r}' \quad \forall r\in \mathbb Z+a, \nonumber \\
\bar{\psi}_{r'}'=&\; ie^{-i\beta}\bar{\tilde{\psi}}_{-r'}' \quad \forall r'\in \mathbb Z-a. 
\end{align}
An incoming boundary state that solves the above gluing condition is
\begin{align}
&|B_{\beta}\rangle_{a} \nonumber\\
&=\exp\left(i e^{-i\beta}\sum_{r'\geq 0}{\psi_{-r'}'\bar{\tilde{\psi}}_{-r'}'}+ie^{i\beta}\sum_{r>0}{\bar{\psi}_{-r}'\tilde{\psi_{-r}'}}\right)|0\rangle_{a,-a},
\end{align} 
while the outgoing boundary state is 
\begin{align}
&_a\langle B_{\beta}| \nonumber\\
&=_{a,-a}\langle0|\exp\left(i e^{i\beta}\sum_{r'\geq 0}{\tilde{\psi}_{r'}'\bar{\psi}_{r'}'}+ie^{-i\beta}\sum_{r>0}{\bar{\tilde{\psi}}_{r}'\psi_{r}'}\right).
\end{align}
Then the closed-channel partition function on the cylinder with $(\beta_{\mathrm{space}}, \ell_{\mathrm{time}})=(2\pi, \pi L)$ is given by
\begin{align}
_a\langle B_{0}| e^{-\pi L H_c}|B_{\beta}\rangle_{a}
=\frac{\vartheta
\left[
 \begin{smallmatrix}
\beta/2\pi - 1/2 \\ -(a-1/2)
\end{smallmatrix}
\right]
(0,iL^{-1})}{\eta(iL^{-1})}.
\end{align}
Identifying $T=L^{-1}$, we find that $|B_{\beta}\rangle_{a}$ is indeed a Cardy state that satisfies the Cardy condition
\begin{align}
Z_{0\beta}^a(T)=_a\langle B_{0}| e^{-\pi L H_c}|B_{\beta}\rangle_{a}.
\end{align}

It is clear that the state $|B_{\beta}\rangle_{a}$ is invariant under both U(1)$_C$ and $(\mathcal{CP})'_{-1}$ (corresponding to $\mathcal{T}_{1}$). This is verified by looking at the symmetry action on the modes (deduced from (\ref{CP'_after_rotation}) and (\ref{mode_after_rotation}) at $\tau'=0$):
\begin{align}
U(1)_{C}:&\; \psi'_{r}\to e^{-i\theta}\psi'_{r}, 
\quad \tilde{\psi}'_{r}\to e^{-i\theta}\tilde{\psi}'_{r}
\nonumber \\
(\mathcal{CP})'_{-\eta}:&\; \psi'_{r}\to e^{i\pi /2}\bar{\tilde{\psi}}'_{r}, 
\quad \tilde{\psi}'_{r}\to -\eta e^{i\pi /2}\bar{\psi}'_{r}, \quad \eta=\pm 1.
\end{align}
Note that we have assumed the ground state $|0\rangle_{a,-a}$ is also invariant under all symmetries.
On the other hand,
the state $|B_{\beta}\rangle_{a}$ is only invariant under U(1)$_C$; it is not invariant under
$(\mathcal{CP})'_{+1}$ (corresponding to $\mathcal{T}_{-1}$).

Let us now consider  
two copies of complex fermions $\{\psi_{1, R}, \psi_{1, L}, \psi_{2, R}, \psi_{2, L}\}$.
One can show that it is now possible to construct a U(1)$_C
\rtimes \mathbb{Z}^{\mathcal{T}_{-1}}_2$ symmetric Cardy state.
One considers the following boundary conditions (before $\pi/2$ space-time rotation):
\begin{align} 
\psi_{1, L}=&\; e^{-i\beta_1}\psi_{2, R}, \quad \bar{\psi}_{1, L}=e^{i\beta_1}\bar{\psi}_{2, R},\nonumber \\
\psi_{2, L}=&\; e^{-i\beta_2}\psi_{1, R}, \quad \bar{\psi}_{2, L}=e^{i\beta_2}\bar{\psi}_{1, R},
\label{complexfermionbc_two_copies}
\end{align}
which preserve U(1)$_C \rtimes \mathbb{Z}^{\mathcal{T}_{-1}}_2$ if $\beta_1=\beta_2\pm\pi$.
The corresponding (total) Cardy state is the tensor product of the outgoing boundary states associated with these two boundary conditions,
\begin{align} 
|B_{\beta_1, \beta_2}\rangle_{a} 
= |B_{\beta_1}\rangle_{a}\otimes |B_{\beta_2}\rangle_{a},
\end{align}
which is invariant under both U(1)$_C$ and $(\mathcal{CP})'_{1}$ (corresponding to $\mathcal{T}_{-1}$).

In summary, a single copy (two copies), or in general, any number of copies (an even number of copies) of the theory (\ref{massless_Dirac}) can be consistently formulated, in the presence of U(1)$_C$ and $\mathcal{T}_{1}$ ($\mathcal{T}_{-1}$) symmetries, on a cylinder $\Sigma$.
Therefore, the BCFT approach agrees with the classification of (2 + 1)$d$ fermionic SPT phases with U(1)$_C$ and time-reversal symmetries given by the gappability argument. In fact, there is a correspondence between the form of Cardy boundary states and gapped phases in (1 + 1) dimensions. In the following section, we study theories of multicomponent bosons, which describe (the edges of) more general SPT phases in (2 + 1) dimensions, and will see such correspondence explicitly.

\section{More general SPT phases in (2 + 1)$d$} 
\label{More general SPT phases in (2+1)D}
\subsection{Canonical quantization}
\label{subsec:CaQuant1}
Let us consider the edge of a (2 + 1)$d$ Abelian SPT phase (either fermionic or bosonic ones) described by the $K$-matrix theory of multicomponent compactified boson fields \cite{LuVishwanath2012},
\begin{align}
S=\frac{1}{4\pi}\int{d^{2}x\left[K_{IJ}\partial_{t}\phi^{I}\partial_{x}\phi^{J}-V_{IJ}\partial_{x}\phi^{I}\partial_{x}\phi^{J}\right]},
\label{KLuttinger}
\end{align}
where $K$ is a $2N\times 2N$ integer-valued symmetric matrix and $I,J=1,\dots,2N$. 
We are interested in studying SPT phases, namely, those that have no topological order, hence we will restrict ourselves to theories with $\det K=1$. Moreover, since SPT phases can be adiabatically connected to trivial phases in the absence of symmetry, their edge theories are always non-chiral. $V_{IJ}$ in Eq.\ (\ref{KLuttinger}) is a non-universal positive definite matrix, which does not affect the topological properties of the theory; $\phi^I$ are compact U(1) bosons that satisfy the compactification condition $\phi^{I}\equiv \phi^{I}+2\pi n^{I}, n^I \in \mathbb{Z}$. When put on a cylinder of circumference $2\pi$, they satisfy the commutation relations
\begin{align}
\left[\partial_{x}\phi^{I}(x),\partial_{x}\phi^{J}(x')\right]=\sum_{m\in \mathbb Z}2\pi i (K^{-1})^{IJ}\partial_{x}\delta(x-x'+2\pi m).
\nonumber
\end{align} 
It is more convenient to carry out the quantization in the redefined basis $\varphi^{I}$ which  we define by diagonalizing the $K$ matrix  
as 
\cite{hsieh2014symmetry}
\begin{align}
\label{eqn:BasisTrans1}
{\bf{A}}{\bf{\phi}}&={\bf{\varphi}}, 
\quad {\bf A}^T \eta {\bf A}= K,
\end{align} 
where ${\bf{A}}\in O(2N)$ and $\eta$ is a diagonal matrix with $\pm 1$ on the diagonal. 
To have a non-chiral theory, we assume 
$\eta$ has equal number of $+1$ and $-1$ in its diagonal.
Without loss of generality we assume $\eta=\mbox{diag}(1,\dots,1,-1,\dots,-1)$.
The theory has $N$ copies of nonchiral bosons. 
The action in the $\varphi$ basis takes the form
\begin{align}
\label{eqn:TransAction}
S&=\frac{1}{4\pi}\int{d^{2}x \left[(\partial_{t}\varphi)^{T}{\bf{\eta}}(\partial_{x}\varphi)
-(\partial_x\varphi)^T(\partial_{x}\varphi)
\right]},
\end{align} 
where we have chosen $V$ such that ${\bf{A}} V {\bf{A}}^{T}=\mathbb I_{2N}$. The Hamiltonian and momentum operators are obtained from the action (\ref{eqn:TransAction}) as
\begin{align}
H&= \frac{1}{4\pi}\int{dx\left[(\partial_x\varphi)^T(\partial_x\varphi)\right]}, \nonumber \\
P&=
\frac{1}{4\pi}\int{dx\left[(\partial_x\varphi)^T\eta(\partial_x\varphi)\right]}.  
\end{align}
After basis transformation, the redefined bosons satisfy the compactification condition
\begin{align}
\varphi^I\sim \varphi^I +2\pi \left({\bf{A}}n\right)^I,
\quad 
n^I\in \mathbb{Z},
\end{align}
and the canonical commutation relation
\begin{align}
\left[\partial_{x}\varphi^{I}(x),\partial_{x}\varphi^{J}(x')\right]=
2\pi i (\eta^{-1})^{IJ}\partial_x
\sum_{m\in \mathbb Z}
\delta(x-x'+2\pi m). 
\nonumber
\end{align}
The mode expansion compatible with the equations of motion,
$\partial_{t}\varphi^{I}\eta_{II}-\partial_{x}\varphi_{I}=0$,
and
the compactification conditions
takes the form
\begin{align}
\label{expansion}
\varphi^{I}&=
\varphi_{0}^{I}
+\frac{2\pi}{L}[t+\text{sgn}(\eta^{II})x]a^{I}_0
\nonumber \\
&\quad 
+\frac{1}{\sqrt{2}}\sum_{r\neq 0}
a_{r}^{I}
e^{
-\frac{2\pi ri}{L}[t+\text{sgn}(\eta^{II})x]}.
\end{align}
Since $\left[\varphi_0^I, a_0^{J}\right]=2\pi i \eta^{IJ}$ and $\varphi_0^I\sim \varphi_0^I +2\pi ({\bf{A}}n)^I$,
\begin{align}
\label{eqn:Quantvarphi}
v^{I}\in \left({\bf{A}}m\right)^{I}\mathbb{Z},  
\quad m^I\in \mathbb{Z},
\end{align}
where $v^I$ is the eigenvalue of $a^I_0$.
The mode operators 
obey the following canonical commutation relation:
\begin{align}
\left[a^{I}_{n},a_{m}^{J}\right]=n\delta^{IJ}\delta_{n+m,0}, \quad n,m \neq 0.
\end{align}

\subsection{Ishibashi states}
States that represent a conformal invariant boundary condition 
are called Ishibashi states.
They satisfy 
\begin{align}
[L_{r}-\bar{L}_{-r}]|I\rangle\!\rangle&= 0,
\label{leftrightvirasoro}
\end{align}
where $L_r$ and $\bar{L}_r$ are the holomorphic and antiholomorphic Virasoro generators, respectively.
For the $K$-matrix theory defined in Eq.\ (\ref{eqn:TransAction}), they are given by  
\begin{align}
L_r&=\frac{1}{2}\sum_{n\in\mathbb{Z}}: (a_{r-n,L})^T a_{n,L}:,
\nonumber \\
\bar{L}_r&=\frac{1}{2}\sum_{n\in\mathbb{Z}}: (a_{r-n,R})^T a_{n,R} :,
\end{align}
where  
\begin{align}
a_r^T=&(a_{r,L}, a_{r,R})^T:=(a^1_r,\ldots, a^N_r, 
a^{N+1}_r,\ldots,a^{2N}_r )^T
\end{align}
are operators that appear in the mode expansion (\ref{expansion}).
While the general solution of (\ref{leftrightvirasoro}) is not known, a sufficient condition for it is given by 
\cite{Oshikawa2010a}
\begin{align}
\label{BS_condition_mode}
(a_{r,L}-R a_{-r,R})|{\bf{v}}\rangle\!\rangle=0, \quad \forall r\in\mathbb{Z},
\end{align}
where the matrix $R\in O(N)$ does not depend on $r$. 
Solutions to Eq.\ \eqref{BS_condition_mode} have the form
\begin{align}
\label{eqn:Ishibashi1}
|{\bf{v}} \rangle\!\rangle:=
\exp\left(\sum_{r=1}^{\infty}\frac{1}{r}(a_{-r,L})^T R a_{-r,R}\right) |{\bf{v}}\rangle,
\end{align}
where $|\bf{v}\rangle$ are eigenstates of $a^I_0$ with eigenvalues $v^I$ that are characterized by Eq.\ (\ref{eqn:Quantvarphi}). 
The Ishibashi condition in Eq.\ (\ref{BS_condition_mode}) can be further simplified by a basis transformation, after which $R$ is rotated to be $\pm \mathbf{1}$. 
Let us clarify this point. 
The Ishibashi condition in Eq.\ (\ref{BS_condition_mode}) is equivalent to 
\begin{align}
\label{eqn:BScondfield}
\varphi_L = R \varphi_R.
\end{align}
Now we can choose a ${\bf{B}} \in O(2N)$ to be
\begin{align}
{\bf{B}} =& \left( \begin{matrix}
1 & 0 \\
0 & R
\end{matrix} \right).
\end{align}
If we redefine the boson fields
\begin{align}
\left( \begin{matrix}
\varphi'_L \\
\varphi'_R
\end{matrix} \right) =& \pm {\bf{B}} 
\left( \begin{matrix}
\varphi_L \\
\varphi_R
\end{matrix} \right), \label{eqn:BosonfieldRot1}
\end{align}
then Eq.\ \eqref{eqn:BScondfield} becomes
\begin{align}
\varphi'_{L/R} = \pm \varphi'_{L/R}.
\end{align}
In terms of the mode operators, we have the Ishibashi condition
\begin{align}
\label{eqn:NewIshibashi}
( a'_{r,L} \mp a'_{-r,R}) 
|{\bf{v}} \rangle\! \rangle = 0, \quad \forall r \in \mathbb{Z}.
\end{align}
This basis rotation and Eq.\ (\ref{eqn:BasisTrans1}) can be simultaneously done if we define ${\bf{A}}' = {\bf{B}} {\bf{A}}$. In the following discussion, we assume this has been done. To lighten the notation, we drop the prime on the field and mode operators. 
\\

\subsection{Equivalence between 
the Ishibashi condition and 
Haldane's null vector condition}
\label{sec:IshibashiHaldaneCond}
Haldane's null vector condition 
of $N$ copies of nonchiral compactified massless bosons 
states:
if there is a set of $N$ linearly independent integer vectors $\{ {\bf{l}}_i \}$ satisfying the condition
\begin{align}
{\bf{l}}_i^T K^{-1} {\bf{l}}_j &= 0,
\quad \forall i,j=1,\cdots,N,
\end{align}
then 
we can find a potential
which can gap out the $N$-component boson theory completely.
\cite{LuVishwanath2012}
This condition comes from the locality requirement 
such that all the bosons can be pinned at the minimum values 
in the gapping potentials simultaneously. 
When Haldane's null vector condition is met, 
one can find the gapping potential
\begin{align}
S_{gapping} =& \sum_{ \{{\bf l} \}} c_{{\bf l}} \int dt \; dx \cos{( {\bf l} \cdot {\bf \phi} + \alpha_{\bf l} ) },
\end{align}
where $\{ {\bf l} \}$ is a set of independent gapping vectors.

In this section,
we discuss the equivalence between the Ishibashi condition and Haldane's null condition. 
We will establish their equivalence at the level of Cardy states,
from which the correspondence between Cardy states and gapped phases (from condensation of independent elementary bosons in the language of Ref. \onlinecite{LuVishwanath2012}) is manifest.


We start from the total Cardy state for the $N$-copy boson system
\begin{align}
\label{eqn:BdyStateT}
|B, \{ \alpha_i \} \rangle=&\otimes_{i=1}^{N}|B, \alpha_i \rangle,
\end{align}
where \cite{gaberdielnotes}
\begin{align}
\label{eqn:Cardyith}
&\qquad\qquad|B,\alpha_i \rangle=\frac{1}{2^{1/4}}\sum_{n_i\in \mathbb Z} e^{in_i \alpha_i} |{\bf{v}}_{i}\rangle\! \rangle_{n_i}
\end{align}
(the repeated indices $i$ are not summed over)
is the Cardy state for the $i$th copy of the system and
\begin{align}
\label{eqn:Ishibashiith}
|{\bf{v}}_i\rangle\!\rangle_{n_i}
&= e^{-\sum_{r >0} (1/r) ({\bf{v}}_{i,L}\cdot a_{-r,L}) ({\bf{v}}_{i,R}\cdot a_{-r,R})}
| n_i{\bf{v}}_i\rangle,
\quad n_i\in\mathbb{Z},
\end{align}
is an Ishibashi state satisfying the Ishibashi condition (\ref{eqn:NewIshibashi}).
Here $\{{\bf{v}}_i=({\bf{v}}_{i,L}, {\bf{v}}_{i,R})\ |\ i=1,..., N\ \text{and}\ {\bf{v}}_{i,L}= -{\bf{v}}_{i,R}\}$ is a set of linearly independent $2N$-component vectors that generates, with integer coefficients, all the solutions satisfying both Eqs.\ (\ref{eqn:Quantvarphi}) and (\ref{eqn:NewIshibashi}).

Note that the Cardy state (\ref{eqn:BdyStateT}) satisfies the Cardy condition
automatically, since it is the direct product of decoupled Cardy states, each
one satisfying Cardy condition separately. 
Now we want to rewrite Eq.\ (\ref{eqn:BdyStateT}) in another form 
in which the connection to the gapping potentials satisfying Haldane's null condition is manifest. 
First, let us write the ground state in a coherent state, namely,
\begin{align}
\label{eqn:VacState1}
| n_i{\bf{v}}_i\rangle = e^{i  n_i {\bf{v}}_i\cdot\varphi_0 } |0\rangle_i
= e^{i  n_i {\bf{e}}_i\cdot\phi_0 } |0\rangle_i,
\end{align}
where $|0\rangle_i = |0\rangle_{i,L}\otimes|0\rangle_{i,R}$ 
is the true vacuum associated with the new zero modes ${\bf{v}}_{i,L/R}\cdot
a_{0, L/R}$ 
and $\{ {\bf{e}}_i := {\bf{A}}^{-1}{\bf{v}}_i \}$ is a set of linearly
independent integer vectors 
[by the definition of $\{{\bf{v}}_i \}$ defined in Eq.\ (\ref{eqn:Quantvarphi})].
Plugging Eqs.\ (\ref{eqn:Cardyith})- (\ref{eqn:VacState1}) into Eq.\ (\ref{eqn:BdyStateT}), 
we obtain
\begin{widetext}
\begin{align}
|B,\{ \alpha_i \} \rangle 
&=\otimes^N_{i=1}   \left(\frac{1}{2^{1/4}} \sum_{n_i \in \mathbb{Z}} e^{i n_i \alpha_i} e^{-\sum_{r >0} (1/r) ({\bf{v}}_{i,L}\cdot a_{-r,L}) ({\bf{v}}_{i,R}\cdot a_{-r,R})} e^{i  n_i {\bf{e}}_i\cdot\phi_0}  |0\rangle_i \right) \nonumber \\
&= \frac{1}{2^{N/4-1}}    e^{-\sum^N_{i=1}\sum_{r >0} (1/r) ({\bf{v}}_{i,L}\cdot a_{-r,L}) ({\bf{v}}_{i,R}\cdot a_{-r,R})} \sum_{\{n_i \in \mathbb{Z}\}}\cos{\left[n_i ( {\bf{e}}_i\cdot\phi_0 + \alpha_i )\right]} |0 \rangle_1 \otimes \cdots \otimes |0\rangle_N.
\label{eqn:BdyStateGapPot}
\end{align}
\end{widetext}

Note that the cosine term in the last line of 
Eq.\ (\ref{eqn:BdyStateGapPot}) is nothing but a gapping potential, and the summation is over all the lattice constructed from the elementary or primitive lattice vectors introduced in Ref.\ \onlinecite{LuVishwanath2012}. Then we conclude that in the $N$-boson system, once we have a Cardy state satisfying the Ishibashi condition, Haldane's null vector condition is also implied, since gapping vectors satisfy Haldane's null condition.


Conversely, given a set of $N$ vectors satisfying Haldane's null vector
condition,
we can always find the set of primitive lattice vectors. Let us assume this is
done.
Then we can construct the Cardy state by following Eq.\
(\ref{eqn:BdyStateGapPot}) backward,
from the bottom to the top line.
This state satisfies the Ishibashi condition and Cardy condition manifestly.

\subsection{Symmetry analysis}
\label{sec:SymAnal}
In the following subsections, in order to facilitate the discussion, we use different bases interchangeably. One can easily see their relations from Eqs.\ (\ref{eqn:BasisTrans1}) and (\ref{eqn:BosonfieldRot1}).

We consider an on-site discrete Abelian symmetry group $G$ with the group action of the form
 \begin{align}
 \hat{g}:{\bf{\phi}} \to {\bf{\phi}}+\delta {\bf{\phi}}^{g}, \quad \forall g\in G,
 \end{align} 
 where we assume that $ \delta {\bf \phi}^{g}$ is constant.
 From the mode expansion of $\phi$, we can read off that $\hat{g}$ only acts on the zero mode $\bf{\phi}_0$ ,
 \begin{align}
 \hat{g}:{\bf{\phi}}_0 \to {\bf{\phi}}_0+\delta {\bf{\phi}}^{g}, \quad \forall g\in G,
 \end{align} 
 Hence a complete set of symmetry invariant gapping potentials, related by boundary conditions, is described by 
 \begin{align}
 \hat{g}:({\bf{l}}^{T} {\bf{\phi}}_0+\alpha)\to ({\bf{l}}^{T} {\bf{\phi}} _0+\alpha) \ \text{mod} \  2 \pi \mathbb Z.
 \end{align}
 From the discussion in Sec. \ref{sec:IshibashiHaldaneCond}, if we have a set of symmetry invariant Haldane vectors, we can find a set of decoupled symmetry invariant Ishibashi states, with which we can construct a symmetry invariant Cardy state. We will show it in the following discussion with two examples.

\subsection{Example: $\mathbb Z_2$ symmetric bosonic SPT}
Let us consider the simple case of 
$\mathbb Z_2$ symmetric bosonic SPT phases. 
The edge theory is described by 
\begin{align}
\mathcal L=\frac{1}{4\pi}\left[(\partial_x\phi)^TK(\partial_t\phi)-v(\partial_x\phi)^T(\partial_x\phi)\right],
\end{align} 
where $K=\sigma^x$.
The $\mathbb{Z}_2$ symmetry,
$\mathbb Z_2=\{e,g\}$, 
acts on the $\phi$ fields as 
\begin{align}
\hat{g}: \left(
\begin{array}{c}
\phi_1\\
\phi_2\\
\end{array}
\right) 
\to 
\left(
\begin{array}{c}
\phi_1\\
\phi_2\\
\end{array}
\right)
+
\pi\left(
\begin{array}{c}
1\\
q\\
\end{array}
\right).
\end{align} 
The theory describes a trivial and a nontrivial SPT phases for $q=0,1$ respectively. 

As claimed above, 
for a trivial SPT phase, that is, for which one can find a symmetric gapping potential, there exists a symmetry invariant boundary state. 
The conditions to be satisfied by a set of symmetric gapping vectors $\left\{{\bf{l}}_i\right\}$ are
\begin{align}
\; {\bf{l}}_i^{T}K^{-1} {\bf{l}}_j&=0, \nonumber \\ 
\;\hat{g}({\bf{l}}_i^{T}\phi+\alpha)\hat{g}^{-1}&=({\bf{l}}_i^{T}\phi+\alpha) \; \text{mod } 2\pi \quad \forall i,j.
\label{gappingcondn}
\end{align}   
Since for the present case we only consider a single nonchiral boson, 
we need to find a single gapping vector ${\bf{l}}$. 
In the case for $q=0$ the above conditions are satisfied by ${\bf{l}}=(0,1)^T$. 
Hence the symmetric gapping term is $\cos(\phi_2+\alpha)$. 
In the chiral basis, this corresponds to 
$\cos\left((\varphi_L-\varphi_R)/\sqrt{2}\right)$. 

On the other hand, 
this gapping potential corresponds to the Dirichlet boundary state that has the gluing condition $a_{0,L}-a_{0,R}=0$. 
The Ishibashi state takes the form
\begin{align}
|{\bf{v}}\rangle\!\rangle=&e^{\sum_{r>0}^{\infty}(1/r) a_{-r} \bar{a}_{-r}}|a_{0,L}=a_{0,R}\rangle,
\end{align} 
and the Cardy state is 
\begin{align}
|B,\phi_0 \rangle=&\frac{1}{\mathcal N_D}\sum_{n\in \mathbb Z}e^{in\phi_0} e^{\sum_{r>0}^{\infty}(1/r)a_{-r}\bar{a}_{-r}}|a_{0,L}=a_{0,R}=n\rangle,
\end{align}
where $\phi_0$ specifies the position of the Cardy state with the Dirichlet boundary condition.

On the other hand, 
in the nontrivial case, namely, for $q=1$,
one cannot find a nontrivial symmetric gapping vector as the conditions \eqref{gappingcondn} imply 
that $l^1 l^2=0$ and $l^1+l^2=0\;\text{mod}\;2$.
These cannot be satisfied simultaneously for any nontrivial $\bf{l}$. 

However, we expect \cite{LuVishwanath2012,chen2012symmetry,chen2013symmetry} a $\mathbb Z_2$ classification so that two copies of the above theory must be trivial.  
This double copy is described by  $\phi:=(\phi^1,\phi^2,\phi^3,\phi^4)^T$ and $K=\sigma^x\oplus\sigma^x$. The symmetry action on the two copies is taken to be identical. In order to be $\mathbb Z_2$ symmetric the two gapping vectors must satisfy
\begin{align}
&l^1_i l^2_j+ l^2_i l^1_j + l^3_i l^4_j + l^4_i  l^3_j = 0, \nonumber \\
&\sum^2_{n=1}l^n_i=0  \ \text{mod} \ 2,
\end{align}  
which comes from Eq. (\ref{gappingcondn}). These conditions can be satisfied simultaneously by the following two gapping vectors:
\begin{align}
{\bf{l}}_1=\left(
\begin{array}{c}
1\\
0\\
0\\
1
\end{array}
\right),
\quad {\bf{l}}_2=\left(
\begin{array}{c}
0\\
1\\
-1\\
0
\end{array}
\right).
\end{align}
This choice is not unique, for example an alternate choice of gapping vectors could be
\begin{align}
\tilde{{\bf{l}}}_1=\left(
\begin{array}{c}
0\\
1\\
1\\
0
\end{array}
\right), 
\quad \tilde{{\bf{l}}}_2=\left(
\begin{array}{c}
1\\
0\\
0\\
-1
\end{array}
\right).
\end{align} 
For the set $\left\{ {\bf{l}}\right\}$, the gapping terms are 
\begin{align}
\label{eqn:Gapping1}
  &
    \mathcal L_{\text{gapping}}
    \nonumber \\
    &=
    \lambda\cos\left(\phi^1+\phi^4+\alpha\right)+\lambda' \cos\left(\phi^2-\phi^3+\alpha'\right)
    \nonumber \\
  &=
    \lambda \cos\left(\frac{1}{\sqrt{2}}(\varphi_{1,L}+\varphi_{1,R}+\varphi_{2,L}-\varphi_{2,R})+\alpha\right)
    \nonumber \\
&\quad +\lambda' \cos\left(\frac{1}{\sqrt{2}}(\varphi_{1,L}
  -\varphi_{1,R}-\varphi_{2,L}-\varphi_{2,R})+\alpha'\right)
  \nonumber \\
&= \lambda \cos{(\Phi_{1,L} + \Phi_{1,R} + \alpha)} + \lambda' \cos{(\Phi_{2,L} + \Phi_{2,R} + \alpha')},
\end{align}
where we define basis transformed bosons 
\begin{align}
\label{eqn:Redefinition}
  \Phi_{1,L} :=& \frac{1}{\sqrt{2}} \big( \varphi_{1,L}+\varphi_{2,L} \big),
                 \nonumber \\
  \Phi_{1,R} :=& \frac{1}{\sqrt{2}} \big( \varphi_{1,R}-\varphi_{2,R} \big),
                 \nonumber \\
  \Phi_{2,L} :=& \frac{1}{\sqrt{2}} \big( \varphi_{1,L} - \varphi_{2,L} \big),
                 \nonumber \\
\Phi_{2,R} :=& -\frac{1}{\sqrt{2}} \big( \varphi_{1,R} + \varphi_{2,R} \big).
\end{align}
The mode expansion of $\Phi_i$  is
\begin{align}
\label{eqn:bmodeexpansion}
\Phi_{i,L}&=
\Phi_{i,0,L}+\frac{2\pi}{L}(t+x){b}_{i,0}
+\frac{1}{\sqrt{2}}\sum_{r\neq 0}
b_{i,r}
e^{-(2\pi ir/L)(t+x)},
\nonumber \\
\Phi_{i,R} &= 
\Phi_{i,0,R}+\frac{2\pi}{L}(t-x){\bar{b}}_{i,0}
+\frac{1}{\sqrt{2}}\sum_{r\neq 0}
\bar{b}_{i,r}
e^{-(2\pi ir/L)(t-x)},
\end{align} 
where ${b}_{1,0}=\frac{1}{\sqrt{2}}(a_{1,0,L}+a_{2,0,L})$, $\bar{b}_{1,0} = \frac{1}{\sqrt{2}} (a_{1,0,R}-a_{2,0,R})$, $b_{2,0}= \frac{1}{\sqrt{2}}(a_{1,0,L}-a_{2,0,L})$ and $\bar{b}_{2,0} = -\frac{1}{\sqrt{2}} (a_{1,0,R}+a_{2,0,R})$. The oscillator modes $b_{i,r}$ for the redefined bosons can be written in terms of mode operators in the original basis based on Eq. (\ref{eqn:Redefinition}).

The redefined mode operators satisfy the following commutation relation:
\begin{align}
\left[b_{i,m},b_{j,n}\right]=m\delta_{m+n,0}\delta_{ij},
\end{align}
and there is a similar relation for the right-moving mode operators.

Hence the symmetry invariant Ishibashi states corresponding to gapping vectors ${\bf{l}}_1$ and ${\bf{l}}_2$ can now be written in terms of symmetric bosons $\Phi_{i,L}+\Phi_{i,R}$,
\begin{align}
|{\bf{v}}_i \rangle\!\rangle_{n}&=
e^{-\sum_{r>0} (1/r) b_{i,-r} \bar{b}_{i,-r}}|b_{i,0}=-\bar{b}_{i,0}=n\rangle.
\label{eqn:Ishibashi2}
\end{align}
We note that these Ishibashi states which are essentially Neumann states for $\Phi_i$ are manifestly symmetric as the bosons $\Phi_{i,L}+\Phi_{i,R}$ are symmetric. The Cardy state constructed from the Ishibashi states is
\begin{align}
|B,\left\{ \alpha_i \right\}\rangle&=
\left( \frac{1}{2^{1/4}}\sum_{n_1 \in \mathbb Z}e^{in_1 \alpha_1}|{\bf{v}}_1 \rangle\!\rangle_{n_1} \right) \nonumber \\
& \quad \otimes  \left( \frac{1}{2^{1/4}}\sum_{n_2 \in \mathbb Z}e^{in_2 \alpha_2}|{\bf{v}}_2 \rangle\!\rangle_{n_2} \right).
\end{align}
To show that this satisfies the Cardy condition, we first compute the amplitude. The closed sector Hamiltonian factorizes in $b^i$ basis as
\begin{align}
H_{\text{c}}&=
\sum_{i=1,2}\left[\frac{1}{2}(b^i_0)^2+
\sum_{r>0}b^i_{-r}b^i_{r}+ \sum_{r>0} \bar{b}^i_{-r} \bar{b}^i_r -\frac{c+\bar{c}}{24}\right].
\end{align} 
Note that this Hamiltonian is not the physical Hamiltonian that we started with for the boson system with boundaries, but the Hamiltonian obtained after 
we perform the $S$-transformation between space and time. 
The amplitude decomposes as 
\begin{align}
\mathcal A=&\;\langle B,{\left\{ \alpha_i \right\}}|q^{H_{\text{c}}}|B,{\left\{ \alpha_i \right\}}\rangle \nonumber \\
=&\; \otimes_{i=1,2}\left[\langle B,  \alpha_i  |q^{{H}^i}|B,  \alpha_i  \rangle  \right],
\end{align} 
where $q=\exp(-2\pi L)$
and we have used the decomposition of the Hamiltonian. 
Both the decomposed parts give rise to the following modular function \cite{blumenhagen2009introduction}:
\begin{align}
\langle B, \alpha_i |q^{H^i}|B, \alpha_i \rangle=&\;\frac{1}{\mathcal N_{N}^2}\frac{1}{\eta(2iL)},
\end{align}
which transforms to the open channel partition function under modular $S$
transformation and hence satisfies the Cardy condition. The subscript ``$N$'' stands for Neumann boundary conditions. 

\subsection{Generalization to $\mathbb{Z}_N$ cases}
The discussion on $\mathbb{Z}_2$ symmetric bosonic SPT phases
can be generalized to case of $\mathbb{Z}_N$ symmetry.
As before \cite{LuVishwanath2012} the edge of a $\mathbb Z_N$ symmetric SPT
is described by a $K$-matrix Luttinger liquid with $K=\sigma^{x}$ in Eq.\ (\ref{KLuttinger}). The symmetry acts as
\begin{align}
\hat{g}: \left(
\begin{array}{c}
\phi_1\\
\phi_2\\
\end{array}
\right) 
\to 
\left(
\begin{array}{c}
\phi_1\\
\phi_2\\
\end{array}
\right)
+
\frac{2\pi}{N} \left(
\begin{array}{c}
1\\
q\\
\end{array}
\right),
\end{align} 
where $\hat{g}$ is the generator of $\mathbb{Z}_N$ group. When $q=0$, this corresponds to a trivial SPT phase and $q=1,\ldots,N-1$ corresponds to nontrivial SPT phases. In analogy to the analysis for the $\mathbb Z_2$ case, one cannot find a symmetric gapping vector when $q\neq 0$. This further implies the inability to find a symmetry invariant Cardy state. However, $N$ copies of a nontrivial $\mathbb Z_{N}$ SPT phase can be deformed to a trivial phase, hence we expect to construct a symmetric boundary state for this enlarged theory.   

We consider $K=\oplus_{i=1}^{N}\sigma^x$, namely, $N$ copies of non-chiral bosons. In this case, the $\mathbb{Z}_N$ symmetry transformation is simply copies of the above transformation, namely
\begin{align}
\hat{g}: \left(
\begin{array}{c}
\phi^1_i\\
\phi^2_i\\
\end{array}
\right) 
\to 
\left(
\begin{array}{c}
\phi^1_i\\
\phi^2_i\\
\end{array}
\right)
+
\frac{2\pi}{N} \left(
\begin{array}{c}
1\\
q\\
\end{array}
\right), \quad i=1,\dots,N.
\end{align} 
To completely gap out the system, we need $N$ $\mathbf{l}$ vectors that satisfy
\begin{align}
\label{eqn:ConstraintK}
{\bf{\mathbf{l}}}_i^T K^{-1} {\bf{\mathbf{l}}}_j =&\; 0, 
\nonumber \\
\hat{g} \; {\bf{\mathbf{l}}}_{i}^T \mathbf{\phi} \; \hat{g}^{-1} =&\; {\bf{\mathbf{l}}}_i^T \mathbf{\phi} \;  \text{mod } 2\pi \quad  \forall i,j. 
\end{align}
Equation (\ref{eqn:ConstraintK}) is equivalent to 
\begin{align}
& \sum^N_{\alpha=1} \big( l^{2\alpha}_i l^{2\alpha-1}_j + l^{2\alpha-1}_i l^{2\alpha}_j \big) =\;0, 
\nonumber \\
& \sum^N_{\alpha=1} l^{2\alpha-1}_i + q l^{2\alpha}_i =\; 0 \ \text{mod} \ N \quad \forall i,j.
\end{align}
Here we choose a simple set of $\bf{l}$ vectors,
\begin{align}
\label{eqn:GappingVectorN}
\{ {\bf{l}} \}: {\bf{l}}_1&= (1,0,1,0,\cdots,1,0)^T \nonumber \\
{\bf{l}}_2&= (0,1,0,-1,0,0,\cdots,0,0)^T \nonumber \\
{\bf{l}}_3&= (0,0,0,1,0,-1,0,0,\cdots,0,0)^T \nonumber \\
\vdots \nonumber \\
{\bf{l}}_N&= (0,0,\cdots,0,1,0,-1)^T.
\end{align}
We can check that in this set, the ${\bf{l}}$ vectors are linearly independent. Then following what is done from Eq.\ (\ref{eqn:Gapping1}), we can write down the gapping potential term
\begin{align}
\mathcal{L}^{\{ {\bf{l}} \}}_{gapping} &= \lambda_1 \cos{(\phi^1_1+\phi^1_3+\cdots + \phi_{N}^1 + \alpha_1)} + \cdots \nonumber \\
&= \lambda_1  \cos{ \big(  \Phi_1 + \alpha_1 \big) } + \cdots, \nonumber
\end{align}
where the redefinitions are
\begin{align}
\Phi_1&= \frac{1}{\sqrt{N}} \big( \phi^1_1 + \phi^1_3+ \cdots + \phi^1_{N} \big) \nonumber \\
\Phi_2 &= \frac{1}{\sqrt{2}} \big( \phi^2_1-\phi^2_2 \big) \nonumber \\
\vdots & \nonumber \\
\Phi_N &= \frac{1}{\sqrt{2}} \big( \phi^2_{N-1} - \phi^2_N \big),
\end{align}
based on the gapping vectors in Eq.\ (\ref{eqn:GappingVectorN}).
Then the $\Phi_i$ fields can be expanded in terms of $b$ fields like those in
Eq.\ (\ref{eqn:bmodeexpansion}).
Then the analysis of Cardy states and the amplitude between boundary states follow that of the $\mathbb{Z}_2$ case. 

We work in the ``$\Phi_i$-bosonic'' basis. In this basis, the Ishibashi states are taken as Neumann free boson states and are manifestly $\mathbb Z_{N}$ symmetric. They take the form
\begin{align}
|{\bf{v}}_i\rangle\!\rangle_{n}=\frac{1}{\mathcal{N}^i_N}\exp\left\{-\sum_{r >0} \frac{1}{r} b_{i,-r} \bar{b}_{i,-r}\right\}|b_{i,0}=-\bar{b}_{i,0} =n\rangle,
\end{align}
where $b_{i,r}$ and $\bar{b}_{i,r}$ are left and right mode operators corresponding to the boson $\Phi_i$ and $b_{i,0}$,$\bar{b}_{i,0}$ are defined similarly as those in Eq.\ (\ref{eqn:Ishibashi2}). The Cardy state for $\Phi_i$ takes the form 
\begin{align}
|B,\alpha_i \rangle=\frac{1}{2^{1/4}}\sum_{n\in \mathbb Z} e^{in \alpha_i} |{\bf{v}}_{i}\rangle\! \rangle_{n}.
\end{align}
The complete boundary state is a tensor product of individual boundary states corresponding to vectors in $\left\{ {\bf{l}} \right\}$,
\begin{align}
|B, \{ \alpha_i \} \rangle=&\otimes_{i=1}^{N}|B, \alpha_i \rangle.
\end{align}

\subsection{General symmetry groups}

Finally, let us consider a more generic symmetry group $G$
acting on the boson fields:
\begin{align}
\hat{g} : \varphi \to U_g \varphi + \delta \varphi^g, \quad \forall g \in G.
\end{align}
The mode operators transform under $g\in G$ as
\begin{align}
\hat{g}: a_r \to U_g a_r,
\quad 
 \varphi_0 \to U_g \varphi_0 + \delta \varphi^g.
\end{align}
In this case, we need to consider both the zero mode part and the oscillator part in Eq.\ (\ref{eqn:BdyStateGapPot}). 
In the previous discussion, we had $U_g = \mathbb{I}$. 
Thus we could focus on the zero mode part, namely, the gapping potential of Eq.\ (\ref{eqn:BdyStateGapPot}). 
To simplify the discussion, we take one copy of compactified boson fields. In this case, $\eta = \sigma^z$ and the mass matrix coupling the left and right moving mode operators can be taken as $M = \sigma^x$ from Eq.\ (\ref{eqn:NewIshibashi}). 
Then the invariance of the Hamiltonian or the action of the theory gives the constraints
\begin{align}
U^T_g U_g &= \mathbb{I}_2, \quad U^T_g \sigma^z U_g = \pm \sigma^z.
\end{align}
Then we have the following general solutions $U_g = \sigma^x, i\sigma^y,
\sigma^z$. When $U_g = \sigma^x$, we have $U_g^T M U_g = M$, which means that
the oscillator part in Eq.\ (\ref{eqn:BdyStateGapPot}) is invariant. However,
when $U_g = i \sigma^y$ or $\sigma^z$, we have $U_g^T M U_g = - M$, meaning that
the oscillator part would flip sign. Physically, it means that the boundary
state changes into
a Dirichlet boundary state from the Neumann boundary state. It is reminiscent of $T$ duality in string theory.  In this case, the zero mode part is usually not invariant. Therefore, for general symmetry groups, we can focus on the zero mode part, which is equivalent to the gapping potential analysis in Ref.\ \onlinecite{LuVishwanath2012}. We will have more discussions on duality in Sec.\ \ref{sec:Discussion}.


\section{(2+1)D topological superconductor protected by $\mathbb{Z}_2\times\mathbb{Z}_2$ symmetry}
\label{(2+1)D topological superconductor protected by Z2xZ2 symmetry}

From the discussion in the last section, we have seen that the construction of a symmetric boundary state is closely related to finding a gapping potential to gap a given (edge) CFT without spontaneous symmetry breaking.
In this section, we show that there is another way to construct a symmetric
Cardy boundary state by considering
only the fundamental boundary conditions of the free fermions.

An example is the class of (2 + 1)$d$ topological superconductors 
protected by a $\mathbb{Z}_2\times\mathbb{Z}_2$ unitary on-site symmetry.
The classification of these topological superconductors is $\mathbb{Z}_8$ 
\cite{ryu2012interacting,2013NJPh...15f5002Q}.
Again, we consider the edge theories, which can be described by
the $N_f$ copies of real fermion fields in 1 + 1 dimensions.
For $N_f=1$, they are described by 
the action 
\begin{align}
S&=\frac{1}{2\pi}\int {d}^{2}xi\bar{\Psi}\gamma^{\mu}\partial_{\mu}\Psi 
\label{realfermionaction}
\end{align}
Upon picking a Clifford basis where $\gamma^{0}=\sigma^{x}$ and $\gamma^{1}=i\sigma^{y}$ and writing $\Psi=(\psi_{L},\psi_{R})$, one can decompose a Majorana fermion into two Majorana-Weyl fermions. 
This action is invariant under a $\mathbb{Z}_2\times\mathbb{Z}_2$ symmetry group that is generated by the fermion number parity for each chirality.

\subsection{Quantization and boundary states}
\label{sec:Realfermions}

Due to the fermionic nature of the fields, there are two sectors depending on the periodicity of the fields under rotations by $2\pi$. The real fermion could have Ramond sector (R) or antiperiodic, Neveu-Schwarz (NS) sector, boundary conditions along the spatial direction. 
For the closed system, the left and right moving fermion fields are decoupled. We can choose boundary conditions independently for them. Therefore, there are four sectors corresponding to the boundary conditions:
\begin{align}
(L,R) =& \text{(R,R),(R,NS),(NS,R),(NS,NS)}.
\end{align}
The fermionic mode expansion takes the form
\begin{align}
\label{eqn:FermionModeExpan}
\psi_{L}(x,t)&=
\sqrt{\frac{2\pi}{L}}\sum_{r}
\psi_{r}e^{-(2\pi ir(t+x)/L)},
\nonumber \\
\psi_{R}(x,t)&=
\sqrt{\frac{2\pi}{L}}\sum_{r}
\tilde{\psi}_{r}e^{-(2\pi ir(t-x)/L)},
\end{align} 
where the mode operators satisfy $\{\psi_{r},\psi_{r'}\}=\delta_{r+r',0}$, $\{\tilde{\psi}_{r},\tilde{\psi}_{r'}\}=\delta_{r+r',0}$ and $\{\psi_{r},\tilde{\psi}_{r'}\}=0$ and $r\in \mathbb Z(+1/2)$ for the Ramond and Neveu-Schwarz sectors respectively.


\paragraph{Boundary state:}
By varying the action \eqref{realfermionaction} and requiring the boundary variation to vanish,
one can read off the suitable 
boundary conditions to be
$
\left\{\psi_{L}\pm \psi_{R}\right\}|_{x=0}=0.
$
In order to construct the Cardy state, we rotate the space-time cylinder by $\pi/2$ such that the manifold has a temporal boundary. Upon space-time rotation the boundary conditions transform to
$
\left\{\psi_{L}\pm i\psi_{R}\right\}|_{t=0}=0
$
These are the relevant boundary conditions for constructing the Ishibashi and Cardy states. The Ishibashi states satisfy the following gluing conditions:
\begin{align}
\left(\psi_{k}+i\eta\tilde{\psi}_{-k}\right)|\eta\rangle\!\rangle=0,
\label{gluingcondition}
\end{align}
where $\eta=\pm 1$. Since this is a free theory, the solutions to the above gluing condition are known. There are two solutions for each $\eta$ corresponding to the NS-NS and R-R sectors. These Ishibashi states are \cite{gaberdielnotes} 

\begin{align}
\label{eqn:IshibashiFermion}
&|\eta\rangle\!\rangle_{NS\text{-}NS}= e^{-i\eta\sum_{r>0}\psi_{-r}\tilde{\psi}_{-r}}|0\rangle_{NS\text{-}NS}, \nonumber \\
&|\eta\rangle\!\rangle_{R\text{-}R}= e^{-i\eta\sum_{r>0}\psi_{-r}\tilde{\psi}_{-r}}|\eta\rangle_{R\text{-}R},
\end{align}
where $|0\rangle_{NS-NS}$ and $|\eta\rangle_{R-R}$ denote the nondegenerate vacuum in the NS-NS sector and the degenerate ground state associated with the $\eta$ boundary condition in the R-R sector, respectively.

Before moving onto the discussion of topological superconductors we note the crucial fact that unless we can construct a Cardy state with only a single boundary condition (namely, $\eta=+1$ or $-1$) in the NS sector, it is impossible to satisfy the Cardy condition without including both sectors. 
This can be seen by considering
the overlap of real-fermion Ishibashi states \cite{Blumenhagen2013a}, 
\begin{align}
&
{}_{NS}\langle\!\langle \eta|e^{-2\pi L H_{c}}|\eta\rangle\!\rangle_{NS}= \frac{\vartheta_{3}(2iL)}{\eta(2iL)},
\nonumber \\
&
{}_{NS}\langle\!\langle \eta|e^{-2\pi L H_{c}}| -\eta\rangle\!\rangle_{NS}= \frac{\vartheta_{4}(2iL)}{\eta(2iL)},
\nonumber \\
&
{}_{R}\langle\!\langle \eta|e^{-2\pi L H_{c}}|\eta\rangle\!\rangle_{R}=
\frac{\vartheta_{2}(2iL)}{\eta(2iL)},
\nonumber \\
&
{}_{R}\langle\!\langle \eta|e^{-2\pi L H_{c}}|-\eta\rangle\!\rangle_{R}= 0,
\end{align}
where $\vartheta_{2,3,4}$ are the Jacobi $\theta$ functions. 
Under modular $S$ transformation, these modular functions transform as
\begin{align}
  \frac{\vartheta_{3}}{\eta(2iL)}\xrightarrow{L=1/2t}&\; \frac{\vartheta_{3}(it)}{\eta(it)},
                                                               \nonumber \\
  \frac{\vartheta_{4}}{\eta(2iL)}\xrightarrow{L=1/2t}&\; \frac{\vartheta_{2}(it)}{\eta(it)},
                                                               \nonumber \\
\frac{\vartheta_{2}}{\eta(2iL)}\xrightarrow{L=1/2t}&\; \frac{\vartheta_{4}(it)}{\eta(it)}.
\end{align}
One can see that unless one can construct a Cardy state with a single Ishibashi state (either $\eta=+1$ or $-1$), the $S$-transformation mixes the $R$-$R$ and $NS$-$NS$ sectors. 

We define the fermion number parity operators,
$(-1)^F$ and $(-1)^{\tilde{F}}$, 
for the left and right moving fermions, respectively, 
which generate $\mathbb{Z}_2 \times \mathbb{Z}_2$ symmetry. 
By construction, these satisfy the following (anti)commutation relations:
$\{(-1)^{F},\psi_{r}\}= 
\{(-1)^{\tilde{F}},\tilde{\psi}_{r}\}=0$, 
and
$[(-1)^{F},\tilde{\psi}_{r}]=
[(-1)^{\tilde{F}},\psi_{r}]= 0$. 

\paragraph{the NS-NS sector:}
%

It is straightforward to check that the $\mathbb{Z}_2 \times \mathbb{Z}_2$ invariant boundary state in the NS-NS sector is 
\begin{align}
|B \rangle_{NS\text{-}NS}&=
\frac{1}{\sqrt{2}}
\big[|+\rangle\!\rangle_{NS\text{-}NS}-|-\rangle\!\rangle_{NS\text{-}NS}\big],
\end{align}
since we have
\begin{align}
(-1)^{F}|\eta\rangle\!\rangle_{NS\text{-}NS}=(-1)^{\tilde{F}}|\eta\rangle\!\rangle_{NS\text{-}NS} =
- |-\eta\rangle\!\rangle_{NS\text{-}NS},
\end{align}
as the vacuum $|0\rangle_{NS\text{-}NS}$ is the eigenstate of both $(-1)^F$ and $(-1)^{\tilde{F}}$ with the eigenvalue $-1$.
It can be seen that both $\eta=\pm 1$ Ishibashi states are needed to construct a fermion parity invariant boundary state in the NS-NS sector. 
We need to include the R-R sector in order to construct a symmetric Cardy state.

\paragraph{the R-R sector:}
The R-R sector is a bit more subtle because of the presence of zero modes. 
Let us define 
\begin{align}
\label{eqn:Gammapm}
\Gamma_{\pm} := \frac{1}{\sqrt{2}} (\psi_0 \pm i \tilde{\psi}_0),
\end{align}
which satisfies the anticommutation relations 
$\{ \Gamma_+, \Gamma_- \} = 1$ 
and $\{ \Gamma_+, \Gamma_+ \} 
= \{ \Gamma_-, \Gamma_- \} =0$.
In terms of the zero mode operators, the fermion parity operators take the following form
\begin{align}
(-1)^{F}&=\sqrt{2}\psi_{0}= 
\Gamma_{+}+\Gamma_{-},
\nonumber \\
(-1)^{\tilde{F}}&=i\sqrt{2}\tilde{\psi}_{0}= \Gamma_{+}-\Gamma_{-}. 
\end{align}
The vacuum in the two sectors $\eta=\pm$ can be defined as
\begin{align}
|\eta=+\rangle =&\; e^{i\Phi_{+}}|0\rangle, \nonumber \\
|\eta=-\rangle =&\; e^{i\Phi_{-}}
\Gamma_{-}|0\rangle,
\end{align}
where $\Phi_{\pm}$ are arbitrary phase factors. 
It can be shown that a fermion parity invariant Ishibashi state does not exist for a single fermion flavor in the R-R sector and consequently we cannot construct a fermion parity invariant boundary state.

\subsection{Boundary states 
and the $\mathbb Z_{8}$ classification} 
\label{sec:RealfermionZ8}

Having found out that,
for a single copy of fermions, 
it is not possible to construct a Cardy state that preserves the fermion number parity, 
we now proceed to analyze multiple copies of real fermions. 
We will show that for $8n$ copies of fermions,
there exists a fermion number parity conserving Cardy state.
This implies a $\mathbb Z_8$ 
classification of topological superconductors. 
This agrees with results in Refs.\ \onlinecite{2013NJPh...15f5002Q,ryu2012interacting}. 


The boundary condition for $N_{f}$ copies of fermions is
\begin{align}
\psi^M + i \eta \tilde{\psi}^M =0, \quad M=1,\dots,N_f.
\end{align}
More generally, 
we may take $\eta^M$ to be different for different copies.
But since later we will take direct 
a product of Ishibashi states with the same $\eta$ value, it is always possible to transform such boundary conditions to the identical $\eta$ case. There could be mixing between different copies, which is the most general case. We do not discuss it here. 
Since one can already construct an NS-NS Ishibashi state for a single flavor of real fermions, 
we will focus our discussion on the R-R sector. We follow the convention in Ref.\ \onlinecite{diaconescu2000fractional}.

We first assume that $N_f$ is even, namely, $N_f = 2n, n \in \mathbb{Z}$. 
It is convenient to define 
\begin{align}
\Gamma^{a \pm} := \frac{1}{\sqrt{2}} (\psi_{0}^{2a-1} \pm i \psi_{0}^{2a}), \quad a=1,\dots,n,
\end{align}
which satisfy the algebra $\{\Gamma^{a+},\Gamma^{b-} \} = \delta^{ab}$, $\{\Gamma^{a+},\Gamma^{b+}\} = \{\Gamma^{a-},\Gamma^{b-}\} =0$. Then the Ishibashi vacua $|\eta \rangle^0_{RR}$ must satisfy
\begin{align}
( \Gamma^{b-} + i \eta \tilde{\Gamma}^{b-}) |\eta \rangle^0_{RR} = 0. 
\end{align}
The solution to this constraint is given by 
\begin{align}
\label{zero_mode_RR}
|\eta \rangle^0_{RR} = e^{-i\eta \sum_b \Gamma^{b+} \tilde{\Gamma}^{b-}} |0\rangle_{RR}, 
\end{align}
where the Fock vacuum is defined as $\Gamma^{a-} |0 \rangle_{RR} =\tilde{\Gamma}^{a+} |0 \rangle_{RR} =0$. 
Finally,
the Ishibashi state in 
the R-R sector can be written as
\begin{align}
\label{Ishibashi_RR}
|\eta\rangle\!\rangle_{RR} = e^{-i \eta \sum_{r>0} \sum^{N_f}_{M=1} \psi^M_{-r} \tilde{\psi}^M_{-r}}|\eta\rangle^{0}_{RR}.  
\end{align}

By construction, $(-1)^F$ anticommutes with left-moving fermionic modes, but commutes with all other modes, while $(-1)^{\tilde{F}}$ anticommutes with all right-moving fermionic modes, but commutes with all other modes. From the expressions (\ref{zero_mode_RR}) and (\ref{Ishibashi_RR}), we thus have
\begin{align}
\label{fermionparity_on_Ishibashi_1}
(-1)^{F (\tilde{F})} |\eta \rangle \!\rangle_{RR} = 
|-\eta \rangle\! \rangle_{RR},
\end{align}
provided $(-1)^F |0\rangle_{RR}=(-1)^{\tilde{F}} |0\rangle_{RR}=|0\rangle_{RR}$.
On the other hand, the fermion number parity operators can also be represented, in the space of the ground states in the R-R sector,  in terms of the zero mode operators as
\begin{align}
(-1)^F &= \left( \frac{1}{i} \right)^n \prod^n_{a=1} \left( 1- 2 \Gamma^{a+} \Gamma^{a-} \right),
\nonumber \\
(-1)^{\tilde{F}} &= \left( \frac{1}{i} \right)^n \prod^n_{a=1} \left( 1- 2 \tilde{\Gamma}^{a+} \tilde{\Gamma}^{a-} \right). 
\end{align}
Using the above relations, one can show  
\begin{align}
\label{fermionparity_on_Ishibashi_2}
(-1)^{F} |\eta \rangle^0_{RR} = \left(-i \right)^n
|-\eta \rangle^0_{RR},
\ 
(-1)^{\tilde{F}} |\eta \rangle^0_{RR} =  i^n
|-\eta \rangle^0_{RR},
\end{align}
which implies, as the action of $(-1)^{F (\tilde{F})}$ on the non-zero modes is as before,
\begin{align}
\label{fermionparity_on_Ishibashi_2}
(-1)^{F} |\eta \rangle \!\rangle_{RR} = \left(-i \right)^n 
|-\eta \rangle\! \rangle_{RR},
\ 
(-1)^{\tilde{F}} |\eta \rangle \!\rangle_{RR} = i^n 
|-\eta \rangle\! \rangle_{RR}.
\end{align}
Now, it seems there are two different ways of how $(-1)^{F (\tilde{F})}$ acts on the Ishibashi states, namely, Eqs (\ref{fermionparity_on_Ishibashi_1}) and (\ref{fermionparity_on_Ishibashi_2}). To avoid this ambiguity, we must require $n=0 \mod 4$ or $N_f=0 \mod 8$ to have a well-defined fermion number parity for each chirality. 

Therefore, the symmetry invariant boundary state in the R-R sector takes the form  
\begin{align}
\label{boundary_state_RR}
|B \rangle_{RR}&=
\frac{1}{\sqrt{2}}\left\{|+\rangle\!\rangle_{RR}+|-\rangle\!\rangle_{RR}\right\}, 
\quad N_f=0 \mod 8. 
\end{align}
The total Cardy states are now the combination of both the NS-NS and R-R parts
\begin{align}
\label{Cardy_state}
|B \rangle_{\pm}&= \frac{1}{\mathcal{N}} \left( |B \rangle_{NSNS} \pm i|B \rangle_{RR} \right),
\quad N_f=0 \mod 8.
\end{align}
The factor $\pm i$ between the NS-NS and the R-R components are both allowed to satisfy the Cardy condition, which also fixes the normalization factor $\mathcal{N}$.

Finally for odd number of flavors of real fermion, there would always be one singlet, which is not paired up. Thus it is impossible to construct a fermion parity invariant boundary state. Therefore the classification is indeed $\mathbb Z_8$.

\subsection{Boundary conditions, gapping potentials and triality}

So far, we have only discussed the transformation of boundary states under symmetry operation. 
But what would happen to the gapping potential? 
Is it also invariant under symmetry operation? 
Here we would like to clarify two points: (1) if we can find a symmetry invariant boundary state, 
then there should exist a set of boundary conditions that is also invariant under the symmetry transformation; (2) symmetry invariant gapping potentials do not guarantee that the corresponding boundary state is also symmetry invariant.

\paragraph{The case of $N_f=8$}
As we have shown before,
for eight copies of Majorana fermions, 
we can construct a fermion parity invariant boundary state, 
which also satisfies the Cardy condition. 
We will now try to identify the corresponding boundary conditions  
following the triality used in Ref.\onlinecite{maldacena1997majorana}.

The boundary conditions and the fermion representation we are using in the notes are given in the vector representation of SO(8) algebra. 
For this algebra, we know that it has an important property--the triality. In the vector representation, we can bosonize the (complex) fermions as
\begin{equation}
\psi_{\alpha j} = e^{-i \phi_{\alpha j}}, \quad 
\psi^{\dagger \alpha j} = e^{i \phi_{\alpha j}},
\end{equation}
where $\alpha, j= 1,2$. Then under the fermion parity operator $(-1)^F$, the boson fields change as
\begin{equation}
\label{eqn:BosonTrans}
(-1)^F \phi_{\alpha j} (-1)^F = \phi_{\alpha j} + \pi.
\end{equation}
Thus for each individual complex fermion, the boundary condition is not
invariant under this $\mathbb{Z}_2$ symmetry.
Now let us use triality to write the fermions in the spinor $(c)$ representation. In this representation, we use a new set of boson fields to bosonize the (complex) fermions. 
\begin{align}
\phi^{\text{ch}} &= \frac{1}{2} \sum_{\alpha,j=1,2} \phi_{\alpha j},
\nonumber \\
\phi^{\text{sp}} &= \frac{1}{2} \sum_{\alpha,j=1,2} (\sigma^z)^\alpha_\alpha \phi_{\alpha j},
\nonumber \\
\phi^{\text{fl}} &= \frac{1}{2} \sum_{\alpha,j=1,2} (\tau^z)^j_j \phi_{\alpha j},
\nonumber \\
\phi^{\text{X}} &= \frac{1}{2} \sum_{\alpha,j=1,2} (\sigma^z)^\alpha_\alpha \; (\tau^z)^j_j \phi_{\alpha j}. 
\label{eqn:RedefinedBoson}
\end{align}
In this basis, with the transformation (\ref{eqn:BosonTrans}),
we can easily check that 
\begin{equation}
(-1)^F \phi^i (-1)^F = \phi^i \quad \text{mod} \; 2\pi, \quad i = \text{ch,sp,fl,X}.
\end{equation}
A similar analysis can be used for the $(-1)^{\tilde{F}}$ operator.
Then, by adding gapping potentials,  
$\phi^i$ defined in Eq.\ \ (\ref{eqn:RedefinedBoson}) 
can be pinned at their ground state values simultaneously. 
Furthermore,
by fermionizing these bosons to define new 
fermion operators, $C_i, i=\text{ch,sp,fl,X}$ via
\begin{align}
C_i &= e^{-i \phi^i},
\end{align}
the boundary conditions can be expressed in terms of $C_i$, namely, 
\begin{align}
C_{L,ch} = C_{R,ch}, \quad& C_{L,fl} = C_{R,fl}, \nonumber \\
C_{L,sp} = - C_{R,sp}, \quad& C_{L,X} = C^\dagger_{R,X} \quad \text{at the boundary}.
\end{align}
Then it is manifest that these boundary conditions are invariant under the
transformations defined
in Eq.\ \eqref{eqn:BosonTrans}.

\paragraph{The case of $N_f=4$}
Let us make the relation between boundary conditions, boundary states and gapping potentials clear. Given a boundary condition, we can obtain a boundary state as the solution to the boundary condition. In this sense, there is a one-to-one correspondence between boundary conditions and boundary states. On the other hand, different gapping potentials can correspond to the same boundary condition. In terms of gapping vectors, it means that primitive and non-primitive lattice vectors can represent the same boundary condition. In this sense, the correspondence between gapping vectors and boundary conditions or boundary states is many to one. Therefore, the symmetry invariance of a specific set of gapping potentials does not imply the symmetry invariance of the boundary condition or the boundary state. Let us take an example to clarify this point. In Ref.\ \onlinecite{isobe2015theory},  the authors show that for two copies of Dirac fermions, which is equivalent to four copies of Majorana fermions, there exists a set of symmetry invariant gapping potential, which is equivalent to the boundary condition. Specifically, in their language of Dirac fermions, the gapping potentials
\begin{align}
V_1 &\propto \psi^\dagger_{1R} \psi^\dagger_{2R} \psi^{\ }_{2L} \psi^{\ }_{1L} + \text{H.c.}= \cos{(2\theta_1+2 \theta_2)},
\nonumber \\
V_2 &\propto \psi^\dagger_{1R} \psi^\dagger_{2L} \psi^{\ }_{1L} \psi^{\ }_{2R} + \text{H.c.} = \cos{(2\theta_1- 2\theta_2)}
\end{align} 
are invariant under fermion parity projection. Here $\theta_a = \frac{1}{2} \left( \phi_{a,R} - \phi_{a,L} \right), a=1,2$ where $\psi^\dagger_{a,R} \propto e^{i \phi_{a,R}}$ and $\psi^\dagger_{a,L} \propto e^{i \phi_{a,L}}$. 
However, 
the $\mathbb{Z}_2 \times \mathbb{Z}_2$ symmetry is spontaneously broken, namely, the single-particle backscattering terms
do not have vanishing vacuum expectation values (vev),
$
\langle \psi^\dagger_{1R} \psi_{1L} \rangle \neq 0$,
$\langle \psi^\dagger_{2R} \psi_{2L} \rangle \neq 0$. In their language, the boundary condition corresponds to the vev. Even if the gapping potential is symmetry invariant, the vev is not invariant. 
This is consistent with our analysis that there is no fermion parity invariant boundary state for four copies of Majorana fermions.

%


\section{Discussion and Outlook}
\label{sec:Discussion}
We have discussed 
the (1 + 1)$d$ edge theories of (2 + 1)$d$ SPT phases
from the perspective of boundary CFT. 
We argue that,
if a (1 + 1)$d$ CFT
is realized as an edge theory of a (2 + 1)$d$ SPT phase,   
it is not possible to find 
a Cardy boundary state
preserving the symmetry of the SPT phase. 
And vice versa:
when 
it is not possible to find a symmetry-preserving
Cardy boundary state in a (1 + 1)$d$ CFT, 
the CFT must be realized 
as an edge theory of a (2 + 1)$d$ SPT phase.  
In short, boundaries of SPT phases are not ``edgeable,''
and, conversely,
``nonedgeable'' CFTs must be realized as an
edge theory of a bulk theory in one higher dimension.

We also observed that
the edgeablity condition in CFTs 
are naturally related to 
the gappability condition. 
This can be seen most straightforwardly 
if one invokes the identification between 
boundary states and 
gapped ground states 
(states obtained from a CFT by adding a massive perturbation).  
Thus, (in)ability to find a symmetry-preserving 
boundary state means 
(in)ability to find a symmetry-preserving gapped 
state.
In turn, this also provides an alternative point of 
view on the relation between 
BCFT and the modular invariance. 
It should be noted that, in higher-dimensional 
SPT phases, 
the gappability condition is replaced 
by a ``weaker'' condition;
(2 + 1)$d$ boundaries of (3 + 1)$d$
nontrivial SPT phases 
are either ingappable or topologically ordered,
if the symmetry of SPT phases is preserved.
Nevertheless, 
the edgeablity condition is still valid 
even for boundaries of higher-dimensional SPT phases. 
Thus,
the edgeablity condition has some precedence
over the gappability condition in general, 
although they seem equivalent 
in (1 + 1)$d$ edges of bulk (2 + 1)$d$ SPT phases. 

In the following,
let us make a few more comments before closing.

\subsection{Symmetry actions on boundary states}

First,
let us summarize 
the way symmetries act on boundary states in CFTs.
In particular, 
we contrast 
physics of 
(2 + 1)$d$ and 
(1 + 1)$d$ SPT phases. 
Let us consider a CFT 
with a global unitary symmetry $G$
(spatial and time-reversal symmetries may be discussed
in a similar fashion). 
We consider 
conformally invariant boundary states
$\{|B_a\rangle \}$ realized in the CFT,
where $a$ labels the boundary states. 
Then, for a symmetry operation $g\in G$, 
one expects
the following possible behaviors of $\{|B_a\rangle\}$
under $g$:
In the first case,
the action of $g$ on boundary states is given by
\begin{align}
g|B_a\rangle_h = 
\varepsilon_a(g|h) 	
|B_a\rangle_h.
\label{case i}
\end{align}
Here,
$|B_a\rangle_h$
is a boundary state in the sector twisted by 
$h\in G$,
and 
$\varepsilon_a(g|h)$ is a phase factor. 
namely, boundary states are invariant under 
the symmetry, up to a phase factor. 
As claimed 
in Ref.\ \onlinecite{2016arXiv160606402C}, 
this case is relevant to 
the physics of boundaries of 
(1 + 1)$d$ SPT phases. 
In Ref.\ \onlinecite{2016arXiv160606402C}, 
the correspondence between gapped ground 
states of (1 + 1)$d$ SPT phases
and boundary states in CFTs was made. 
These boundary states are anomalous 
in the sense that when acted with symmetry
they give rise to anomalous U(1) phases, Eq. \eqref{case i}.
Furthermore, the anomalous phase $\varepsilon(g|h)$
is related to the two cocycles in $H^2(G,\text{U(1)})$,
which gives the classification of (1 + 1)$d$ SPT phases 
protected by $G$. 
(These phases, however, only appear in 
boundary states in twisted sectors,
namely, the sectors with twisted boundary conditions
by a group element in $G$.)

On the other hand, 
there are cases in which 
a boundary state $|B_a\rangle$ 
is mapped to 
another boundary state
$|B_{a'}\rangle$, 
which can be different from the original one: 
\begin{align}
g|B_a\rangle = |B_{a'}\rangle.
\label{case ii}
\end{align}
We further distinguish the following two cases: 
(a) There is a subset of boundary states which are 
mapped to themselves for all symmetry operations $g\in G$.
(b) None of the boundary states remain invariant under $g\in G$.  
Case (a) is a typical situation when 
the (1 + 1)$d$ CFT can be realized on its own right, without 
referring to higher-dimensional bulk systems. 
On the other hand,
Case (b) is relevant to (2 + 1)$d$ SPT phases,
as we have discussed for the bulk of the paper. 

\subsection{Boundary states and locality}

In Eq.\ \eqref{case ii},
it should be noted that the right-hand side is not given by
a superposition of $|B_a\rangle$, but by a single 
boundary state.
In fact, 	superpositions of $|B_a\rangle$
in general do not satisfy the Cardy condition, and are disqualified
as a physical boundary state.
[In this respect, the symmetry transformation law 
in Eq.\ \eqref{case ii} is analogous to anyonic symmetry
which acts on (2 + 1)$d$ topologically ordered phases by permuting anyons.]
In the present context,
this is in perfect agreement with the standard theory of 
spontaneous symmetry breaking. 
When spontaneous symmetry breaking happens,
ground states having different expectation values 
of an order parameter should not be superposed in the thermodynamic limit. 
(These states are ``superselected.'')
The overlap of these states vanishes in the thermodynamic
limit, and hence a given ground state
with a definite value of the order parameter cannot  
be mixed by any physical (local) operation.
(However, the overlap between different boundary states
may not be zero, and defines the Affleck-Ludwig $g$ function.)

In some sense, 
one can think of Cardy states setting the notion of locality. 
It should be noted that there are 
multiple sets of solution to the Cardy equations, 
which correspond to different modular invariant
bulk partition functions.

%

Let us further illustrate the notion of locality set 
by the Cardy states:
%
As we demonstrated through various examples,  
when none of the boundary states are invariant under 
symmetry $G$,
the CFT must be realized as an edge theory of a bulk 
nontrivial SPT phase
protected by on-site unitary symmetry $G$.
In the edge theory, 
the criticality (gapless spectrum) is 
enforced by the symmetry $G$.  
This is quite different from 
criticalities (conformal field theories)
that occur in isolated (1 + 1)$d$ systems;
there are typically perturbations at a critical point 
which are $G$ symmetric. 
By perturbing the critical point by such perturbation, 
it may be possible to flow into a gapped phase where the 
$G$ symmetry is preserved. 
This suggests that
the symmetry $G$ acting within the edge theory of
a nontrivial SPT phase is not an ordinary symmetry. 
In fact, 
as noted in Ref.\ \onlinecite{santos2014symmetry}, 
the symmetry $G$ is realized non-locally or as a non-on-site 
symmetry within the edge theory.

\subsection{Duality and triality}

Another canonical example is provided by 
the $\mathbb{Z}_2$ symmetric topological superconductor 
discussed in 
Sec. \ref{(2+1)D topological superconductor protected by Z2xZ2 symmetry}. 
The edge theory in this case is described by 
the action \eqref{realfermionaction}. 
Here,
the $\mathbb{Z}_2$ symmetry flips the 
sign of the mass term, and hence enforces the criticality. 
In the language of the (1 + 1)$d$ transverse-field quantum Ising model 
(or the 2$d$ Ising model),
this is nothing but the Kramers-Wannier duality. 
It is a non-local operation which exchanges 
the Ising spin operator $\sigma$ and the disorder operator $\mu$.

Let us have a look at how this $\mathbb{Z}_2$ symmetry acts on 
boundary states. 
In the critical Ising model,
there are three physical conformal boundary
conditions: the free condition $|f\rangle$, 
and the fixed ones $|+\rangle$
and $|-\rangle$. 
The periodic (R) sector contains three scalar fields:
the identity, 
the spin field $\sigma$, and the energy density $\varepsilon$, 
of chiral conformal weight 0, 1/16, and 1/2
respectively. They
lead to three Ishibashi states 
$
|0\rangle\!\rangle_{\mathrm{R}}
$,
$
|\frac{1}{16}\rangle\!\rangle_{\mathrm{R}}
$,
and
$
|\frac{1}{2}\rangle\!\rangle_{\mathrm{R}}
$.
The
second, antiperiodic (NS) sector contains a single scalar field,
the disorder field $\mu$, 
with the same conformal weight 1/16 as
the spin field, and gives rise to one Ishibashi state 
$
|\frac{1}{16}\rangle\!\rangle_{\mathrm{NS}}
$.
The Cardy boundary states are given in terms of 
these Ishibashi states as
\begin{align}
|\pm \rangle 
&=
\frac{1}{\sqrt{2}}
\left[
|0\rangle\!\rangle_{\mathrm{R}}
\pm 
\sqrt{2}^4
|\frac{1}{16}\rangle\!\rangle_{\mathrm{R}}
+
|\frac{1}{2}\rangle\!\rangle_{\mathrm{R}}
\right],
\nonumber \\
|f\rangle 
&=
|0\rangle\!\rangle_{\mathrm{R}}
+
\sqrt{2}^4
|\frac{1}{16}\rangle\!\rangle_{\mathrm{NS}}
-
|\frac{1}{2}\rangle\!\rangle_{\mathrm{R}},
\end{align}
which, in terms of the Ising spin variables, correspond 
to 
the fixed boundary condition with spin pointing up/down at the boundary,
and the free boundary condition.
The Kramers-Wannier duality exchanges
the free boundary condition
$|f\rangle$
and 
one of the fixed boundary conditions ($|+\rangle$). 
This is so since the duality transformation exchanges 
$\sigma$ and $\mu$, and hence 
the Ishibashi states
$|1/16\rangle\!\rangle_{\mathrm{R}}$
and
$|1/16\rangle\!\rangle_{\mathrm{NS}}$.
(In fact, 
Ref.\ \onlinecite{Ruelle:2005nk} proposed a method 
to diagnose the existence of the Kramers-Wannier duality, 
for a given CFT, by using boundary states.)

Let us next consider 
$N_f$ copies of 
(2 + 1)$d$ topological superconductors protected by 
$\mathbb{Z}_2$ symmetry, as
discussed.
We will focus on the cases where $N_f$ is even. 
In these cases, 
spin operators (analog of $\sigma$ and $\mu$ in
the critical Ising model) 
in the edge theory are given by 
\begin{align}
 \Theta^{\boldsymbol{s}}_R
 =
 e^{ i \sum_a s_a \phi^a_R},
 \quad
 s_a = \pm \frac{1}{2}
\end{align}
in the bosonized language (in the right-moving sector). 
These operators are
an intertwining (vertex) operator that maps 
the untwisted sector to the twisted sectors specified by $\boldsymbol{s}$. 
By state-operator correspondence, these operators are identified with 
a state in the corresponding twisted sector. 
(Note that there is ground-state degeneracy for the R sectors.) 
%
%
Thus, we have a set of states 
$
 \{ |0 \rangle_{\mathrm{NS}}, |\boldsymbol{s}\rangle_{\mathrm{R}} \}.
$
These states appear when one constructs boundary states,
and are exchanged under the action of 
the unitary $\mathbb{Z}_2$ symmetry. 
(Here, this is not the $\mathbb{Z}_2^f$ symmetry.)  
%
The spin operators satisfy 
\begin{align}
 \Theta_R^{\boldsymbol{s}} (z) 
 \Theta_R^{\boldsymbol{s}'}(w)
&= 
e^{2\pi i \boldsymbol{s}\cdot 
\boldsymbol{s}' \mathrm{sgn}(x-x') }
 \Theta_R^{\boldsymbol{s}'}(w)
 \Theta_R^{\boldsymbol{s}} (z),
\end{align}
where 
$
 \boldsymbol{s} \cdot \boldsymbol{s}' =
 (1/4)\sum_a (\pm 1)
$.
This phase can be made an integer when
the number of complex fermions is a multiple of 4
(namely, the number of real fermions is 
a multiple of 8, $N_f = 8n$)
and if we choose
\begin{align}
 \boldsymbol{s} = (1/2, 1/2, \cdots)
 \quad
 \mbox{or}
 \quad
 \boldsymbol{s}'=(-1/2, -1/2,\cdots). 
\end{align}
The unitary $\mathbb{Z}_2$ symmetry can exchange
spin operators $\Theta_R^{\boldsymbol{s}}$,
as 
the Kramers-Wannier duality of the critical Ising model
exchanges $\sigma$ and $\mu$.
However when $N_f=8n$, 
the spin operators are mutually local.
Hence in this case, 
the $\mathbb{Z}_2$ symmetry is not a duality 
(or non-local) symmetry. 
Rather, it is a (part of) triality symmetry.

%
%
%
%
Finally, 
recall that
the presence of boundary breaks the 
Kramers-Wannier duality.
This is another indication that 
the Kramers-Wannier duality 
is nonlocal.
	
%

\acknowledgements

We thank T.Hughes, R.Leigh and K.Shiozaki for useful discussions.
This work was supported in part 
by National Science Foundation Grant No. DMR-1455296.

\appendix
\bibliography{bcftreference}

\end{document}